\documentclass[twocolumn,showpacs,preprintnumbers,amsmath,amssymb]{revtex4}
\usepackage{amsmath,amsfonts,latexsym,amssymb,graphics,epsfig,subfigure,color,makeidx}
\usepackage{xcolor}
\usepackage{multirow}
\newcommand {\nn}    {\nonumber}

\begin{document}

\title{Localization and mass spectra of various matter fields on Weyl thin brane}

\author{Tao-Tao Sui\footnote{suitt14@lzu.edu.cn}$^{a}$,
                 Li Zhao\footnote{lizhao@lzu.edu.cn, Corresponding author}$^{a}$,
              Yu-Peng Zhang\footnote{zhangyupeng14@lzu.edu.cn}$^{a}$,
              Qun-Ying Xie\footnote{xieqy@lzu.edu.cn}$^{b}$
}
\affiliation{
                 $^{a}$Institute of Theoretical Physics, Lanzhou University, Lanzhou 730000, People's Republic of China\\
                 $^{b}$School of Information Science and Engineering, Lanzhou University, Lanzhou 730000,
                  People's Republic of China}

\begin{abstract}
It has been shown that the thin brane model
in a five-dimensional Weyl gravity can deal with the wrong-signed Friedmann-like equation in the Randall--Sundrum-1 (RS1) model. In the Weyl brane model, there are also two branes with opposite brane tensions, but the four-dimensional graviton (the gravity zero mode) is localized near the negative tension brane, while our four-dimensional universe is localized on the positive tension brane.
In this paper, we consider the mass spectra of various bulk matter fields (i.e., scalar, vector, and fermion fields) on the Weyl brane. It is shown that the zero modes of those matter fields can be localized on the positive tension brane under some conditions. The mass spectra of the bulk matter fields are equidistant for the higher excited states, and relatively sparse for the lower excited states. The size of the extra dimension determines the gap of the mass spectra. We also consider the correction to the Newtonian potential in this model and it is proportional to $1/r^{3}$.
\end{abstract}


\maketitle

\section{Introduction}\label{scheme1}

Over the last decades, there was a great interest to construct a unified theory in
higher-dimensional spacetime. The higher-dimensional theory dates back to the work of Kaluza
and Klein \cite{KK26} in the 1920s. They tried to unify electromagnetism with gravity by assuming that the $g_{5\mu}$ component of the five-dimensional metric represents the electromagnetic potential \cite{KK26}. Subsequently, the brane world theories appear as an alternative theory to solve the gauge hierarchy and cosmological constant problems (see e.g., Refs. \cite{Rubakov,RSmodel1,Randall2,Arkani-Hamed1998}) with a fundamental idea that the visible
universe is localized on a 3-brane embedded in a higher-dimensional bulk.

Most research  as  regards about brane worlds is based on a Riemann geometry
\cite{Bajc2000,Fuchune,Hatanaka1999,Das2008,Volkas2007,RandjbarPLB2000,Guo1103,Liu2009uca,
Shiromizu2000}. The non-Riemannian models can be classified within the language of metric-affine gravity \cite{Liu0708,Puetzfeld2005,Hehl1995,Hammond2002}.
Generally, there are two types of generalized Riemann geometries, Riemann--Cartan geometry
and Weyl geometry~\cite{Puetzfeld2005}.
Cartan introduced the torsion $T$ to represent the antisymmetric
piece of the connection, which may be related to spin. When the connection includes a geometric scalar
$\omega$, there will be a nonmetricity part called $Q$, which is related to the scalar $\omega$.
In Riemann--Cartan geometry, the torsion $T\neq0$, and the nonmetricity part $Q=0$.
When $T=0$, the Riemann--Cartan geometry can degenerate into Riemann geometry.
The Weyl geometry is an affine manifold described by a pair ($g_{MN}$, $\omega_{M}$), where $g_{MN}$ is the metric and $\omega_{M}$ is the ``gauge" vector as the gradient of a scalar function $\omega$. In the Weyl geometry, $T=0$ and $Q\neq0$. When $Q=0$, the Weyl geometry can also degenerate into Riemann geometry. As a generalization of Riemann, Weyl geometry has been received much attention in the studies in field theory and cosmology
\cite{Barbosa2006,ChengPRL1998,Hochberg1991,MoonJCAP2010}.
Moreover, in Refs.
\cite{Barbosa2008,LiuJHEP2008,Correa1607,Arias0202,Cendejas0603,Liu0708,Liu0911},
the authors have studied the Weyl brane world scenario and obtained various solutions of Minkowski thick branes.

In the brane world scenario with all matter fields (scalar, vector, and fermion fields) propagating in the bulk, the zero modes of these bulk fields, i.e., the four-dimensional matter fields in the Standard Model, should be localized on the brane complying with the present observations. So, localization of various bulk
matter fields on a brane is an important and interesting issue.
References~\cite{Bajc2000,Fuchune,Furlong1407} showed that the zero mode of a free massless scalar field can be localized on the RS brane or its generalized branes.
{In order to ensure that the zero mode of vector field can be localized on the brane, the case of generalized RS branes with more extra dimensions was considered  \cite{Oda2000,Sakamura1607}, a dynamical mass term was included in gauge field localization \cite{zhao1406,alencer1409}, and in Refs. \cite{Araujo1406,Costa1304, Aguilar1401,Costa1501,sui1703,Chumbes1108,Cruz1211,Zhao1402,Alencar1005,Alencar1008} the authors introduced the interaction with the background scalar field}.
Localization of the fermion zero mode on a brane is especially important. It has been demonstrated that localization mechanisms, e.g. Yukawa  coupling $\eta\bar{\Psi}F(\phi)\Psi$ \cite{Liu0708,Volkas2007,RandjbarPLB2000,Guo1103,
Fuentevilla1412,Cendejas1503,Rocha1606,Brito1609,KoleyCQG2005,
D. Bazeia0809,MelfoPRD2006,Almeida0901,Almeida2009,Davies0705,LBCastro1,LBCastro2},
derivative scalar¨Cfermion coupling $\eta\bar{\Psi}\Gamma^{M}\partial_{M}F(\phi)\gamma^{5}\Psi$
\cite{Liu1312}, and derivative geometrical--fermion coupling\\
$\eta\bar{\Psi}\Gamma^{M}\partial_{M}F(R)\gamma^{5}\Psi$  \cite{Liu1701}, can ensure that the fermion
zero mode can be localized on the brane.

In the RS1 model \cite{RSmodel1}, the authors assumed that there are two thin branes located at the boundaries {of} the
extra dimension. One of them is the negative tension brane (called Tev brane, where our
universe is resided), and the other is the positive tension brane (called the Planck brane, where the
spin-2 gravitons are localized). The warping of the extra dimension in RS1 model can solve the famous
gauge hierarchy problem, but the fundamental assumption that the our universe is localized on the negative
tension brane can lead to a wrong-signed Friedmann-like equation \cite{ Csaki99,Cline99,Shiromizu99}.
The authors of Ref.~\cite{Yang:2011pd} presented a generalized RS1 model in the Weyl integrable geometry, which can give a correct Friedmann-like equation since our universe is assumed to be located on the positive tension brane. It was shown that the brane system is stable under the tensor perturbation and the gravity zero mode is localized near the negative tension brane, whereas the matter fields should be confined on the positive tension brane, since our universe is located on this brane.

In this paper, we will  do further research on the localization of various bulk matter fields in this Weyl brane model and check whether the zero modes of these fields can be confined on the positive tension brane.
We will consider a dilaton coupling between the bulk matter fields and
the background scalar field. It will be found that the coupling is necessary for most of the bulk fields.
For the fermion case, we consider the usual Yukawa coupling $\eta\bar{\Psi}F(\omega)\Psi$ to obtain analytical
wave functions of the fermion KK modes. This paper is organized as follows. In Sect.~\ref{STbrane_SecModel}, we give a brief review of the Weyl brane model proposed in Ref.~\cite{Yang:2011pd}. In Sect.~\ref{STbrane_SecBulkMatterFieldLocalize}, we investigate localization and mass spectra of the
scalar, vector, and fermion fields on the Weyl brane. In Sect.~\ref{gravitational perturbations}, gravitational
fluctuations and correction to the Newtonian potential are discussed. Finally, we give a brief conclusion and summary in Sect.~\ref{STbrane_secConclusion}.

\section{Review of Weyl brane model}\label{STbrane_SecModel}

The brane model we consider here is based on the  five-dimensional Weyl geometric action where the
gravity is non-minimally coupled with a background scalar field $\omega$. The action is given by
\cite{Yang:2011pd}
\begin{equation}
S_5=\frac{M^{3}_{*}}{2}\int_{M_{5}^{W}}{d^5x\sqrt{-g}e^{k\omega}
\left[R-\left(4+4k\right) \nabla_K \omega \nabla^K \omega \right]}.\\
\label{STbrane_action}
\end{equation}
Here $R$ is the five-dimensional Weyl Ricci scalar, $M^{-3}_{*}\equiv8\pi G_5$ is the fundamental scale of
gravity, the Latin letters $M,N,K =0,1,2,3,5$, and $\omega_{,K}$ is a gradient of the Weyl scalar $\omega$. In what follows it is convenient to denote  the physical four coordinates by $x^\mu$, and the extra dimension  by $y$.

In this frame the Weyl Ricci tensors are constructed from $\Gamma_{LN}^{M}$ in  a  standard  way,
\begin{eqnarray}
R^{M}_{LKN}&=&\partial_{K}\Gamma_{LN}^{M}-\partial_{N}\Gamma_{LK}^{M}
  +\Gamma_{KP}^{M}\Gamma^{P}_{NL}-\Gamma_{NP}^{M}\Gamma^{P}_{KL}, \nonumber\\
R_{LN}&=&R^{M}_{LMN},~~~ \\
R&=&R_{5}=R^{L}_{L}, \nonumber
\label{STbrane_Riemann }
\end{eqnarray}
where the Weyl affine connection is expressed as
\begin{eqnarray}
\Gamma^{P}_{MN}=\{^{P}_{MN}\}
  -\frac{1}{2}\left(\delta^{P}_{N}\nabla_M \omega+ \delta^{P}_{M} \nabla_N \omega-
g_{MN} \nabla_P \omega\right),
\label{STbrane_gamma}
\end{eqnarray}
with $\{^{P}_{MN}\}$ representing the Riemannian Christoffel symbol. Now with some algebra,
the Weyl Ricci tensor $R_{MN}$ can be separated into a Riemannian Ricci tensor
part $\hat{R}_{MN}$ plus some other terms with respect to the Weyl scalar function,
i.e.,
\begin{eqnarray}
R_{MN} &=&\hat{R}_{MN}+\frac{3}{2} \nabla_M \nabla_N \omega+\frac{1}{2}g_{MN} \nabla_P \nabla^{P} \omega \nonumber\\
                 &&+\frac{3}{4}(\nabla_M \omega \nabla_N \omega-g_{MN} \nabla_K \omega \nabla^K \omega),
\end{eqnarray}
where the hatted magnitudes and operators are defined by the Christoffel symbol.
 Then from the action (\ref{STbrane_action}), the equations of motion are
\begin{eqnarray}
   \hat G_{MN}&=&(4+4k)\Big(\nabla_{M}\omega
   \nabla_{N}\omega-\frac{1}{2}g_{MN}\nabla^{K}\omega\nabla_{K}\omega\Big)\nonumber\\
            &&+k\Big(\nabla_M\nabla_N\omega-g_{MN}\nabla^K\nabla_K \omega \Big)\nonumber\\
            &&+k^2\Big(\nabla_M\nabla_N\omega-g_{MN}\nabla^K\omega\nabla_K \omega\Big),
            \label{EOM1}\\
k\hat R&=&-(8+8k)\nabla^K\nabla_K\omega-4(k+k^2)\nabla^{K}\omega\nabla_{K}\omega.\label{EOM_2}
\end{eqnarray}

Further, we can write the field equation (\ref{EOM1}) in the form of  the Einstein equation by moving all Weyl
scalar terms to the right-hand side to compose an effective energy-momentum tensor:
\begin{equation}
 \hat G_{MN}=\hat T_{MN}.\label{Efective_EEq}
\end{equation}
The authors  in Refs.  \cite{Barbosa2006, Barbosa2005, Barbosa2008,Vladimir0904,Vladimir0912,
D. Bazeia1502,Marzieh1504} studied thick brane solutions, by considering the metric that satisfies
four-dimensional Poincar$\acute{e}$ invariance. The localization and mass spectrum problems of bulk
matter fields on the Weyl Minkowski thick branes were discussed in Refs.
\cite{Barbosa2005,Barbosa2008,LiuJHEP2008,Correa1607,Arias0202,Cendejas0603,Liu0708,Liu0911}.
We are interested in the Weyl Minkowski thin branes with space $S^1/Z_2$, for which the metric with
four-dimensional Lorentz invariance in a five-dimensional Weyl spacetime is given by
\begin{eqnarray}
ds^2 &=& a^2(y) \eta_{\mu\nu}dx^\mu dx^\nu+ dy^2 \nonumber\\
         &=&a^2(z)\left(\eta_{\mu\nu}dx^\mu dx^\nu+ dz^2 \right),  \label{STbrane_linee}
\end{eqnarray}
where  $z$ is the conformal coordinate of the extra dimension with $z \in [-z_b, z_b]$
and it relates to the physical coordinate $y$ by a coordinate transformation $dy = a(z)dz$. One assumes that 
the background scalar field $\omega$ only depends on the extra dimension.

According to the metric (\ref{STbrane_linee}), the field equations in the bulk can be expressed in terms of
 the warp factor $a$, and  the scalar field $\omega$,as in the  following set  of equations:
\begin{subequations}\label{Metric_EOM_2}
\begin{eqnarray}
k\omega''+(k^2+2k+2)\omega'^2 \nonumber\\
+2k\frac{a'}{a}\omega'+3\frac{a''}{a} &=& 0, ~\label{Metric_EOM_2_1}\\
    (1+k)\omega'^2-2k\frac{a'}{a}\omega'-3\frac{a'^2}{a^2}&=&0,\label{Metric_EOM_2_2}\\
2(1+k)(\omega''+3\frac{a'}{a}\omega')+k(1+k)\omega'^2\nonumber\\
-k(\frac{a'^2}{a^2}+2\frac{a''}{a})&=&0,
    \label{Metric_EOM_2_3}
\end{eqnarray}
\end{subequations}
where the prime denotes the derivative with respect to $z$. The above field equations admit the following
brane solution \cite{Yang:2011pd}:
\begin{subequations}\label{BraneSolution}
\begin{eqnarray}
a(z)&=&\left( {1 + \beta |z|} \right)^{\frac{1}{3} (1+\frac{k}{\alpha})},\label{Sol_WP}\\
\omega(z)&=&-\frac{1}{\alpha}{\ln({1+\beta|z|})},\label{Sol_Sc}
\end{eqnarray}
\end{subequations}
where $\alpha=\sqrt{k^2+3k+3}$, the parameter $\beta > 0$ and $k<-1$.\\
The five-dimensional effective energy-momentum tensor is given as usual by
\begin{eqnarray}
\hat{T}_{MN}=\frac{2}{\sqrt{-g}}\frac{\delta \hat{S}}{\delta g_{MN}}=\hat{R}_{MN}-\frac{1}{2}g_{MN}\hat{R},
\end{eqnarray}
with $\hat{S}$ the effective matter action. In this case the effective energy density is given  by the
null-null component $\hat{T}_{0}^{0}$ of the energy-momentum tensor,
\begin{eqnarray}
\rho(z)&=&\frac{\beta^2}{3\alpha^2}\frac{(k^2+\alpha k+6k+6)}{(1+z\beta)^{\frac{2}{3}(4+\frac{k}{\alpha})}} \nonumber\\
       &&-\frac{2k\beta[\delta(z)-\delta(z-z_b)]}{\alpha(1+\beta|z|)^{\frac{5}{2}+\frac{2k}{3\alpha}}}.
\end{eqnarray}

This expression directly shows that  there are  two delta functions appearing at each boundary, so
they stand for two thin branes, namely a positive tension brane at the origin and a negative one at the
boundary $z_{b}$.  These tensions compensate for the effects produced by the bulk component and
hence ensure the existence of four-dimensional flat branes. So the brane configuration of this case is
similar to that of the RS1 model.  However,  localization of massless graviton in these two brane
scenarios  will be quite different, and in Ref. \cite{Yang:2011pd} one can  suppose that the standard model fields are confined on the positive tension brane at $z=0$, so this is crucial to overcome the severe cosmological problem of the RS1 model.

\section{Localization of various matter fields}
\label{STbrane_SecBulkMatterFieldLocalize}
In this section, we investigate the localization and mass spectra of various bulk matter fields, including the spin-0 scalar, spin-1 vector and spin-1/2 fermion fields on the Weyl brane. A  natural  mechanism for  localizing spin-0 scalar and  spin-1 vector fields on the brane  involves the  couplings with  the background scalar field. For the case of spin-1/2 fermions, we will investigate the usual Yukawa coupling $\eta \bar \Psi F(\omega)\Psi$.

\subsection{Spin-0 scalar field}
\label{STbrane_sec3.1}

Firstly,  we explore the localization condition and the mass spectrum for a massless scalar field on the Weyl brane by considering two types of coupling with the background scalar fields, dilaton coupling and Higgs potential coupling \cite{Liu2009uca}.

\subsubsection{The dilaton coupling}
\label{STbrane_sec3.1.1}
Let us consider a massless scalar field $\Phi$ coupled with the dilaton field $\omega$ via the following five-dimensional action:
\begin{equation}
S_{\text{s}} = -\frac{1}{2}\int d^5 x  \sqrt{-g}~ e^{\lambda\omega} g^{MN} \partial_M \Phi \partial_N\Phi,
\label{STbrane_scalar_action}
\end{equation}
where $\lambda$ is a dimensionless coupling constant. For convenience, we rewrite the warp factor so that  the metric (\ref{STbrane_linee}) takes the form
\begin{eqnarray}
ds^2 &=& a^2(z)\left( \eta_{\mu\nu}dx^\mu dx^\nu+ dz^2 \right)\nonumber\\
&=& e^{2A(z)}\left( \eta_{\mu\nu}dx^\mu dx^\nu+ dz^2 \right).\label{STbrane_linee1}
\end{eqnarray}
Following (\ref{STbrane_scalar_action}) and (\ref{STbrane_linee1}), the equation of motion for the scalar field satisfies the five-dimensional Klein¨CGordon equation,
\begin{eqnarray}
 \partial_{P}\left(\sqrt {-g}\partial^{P} \Phi \right)=0.\label{EOM of scalar field}
\end{eqnarray}
Making the KK decomposition
\begin{eqnarray}
 \Phi(x^{\mu},z)=\sum_n \phi_n(x^\mu) \chi_n(z) e^{-\varrho A(z)},
 \label{kk decompotition}
\end{eqnarray}
with the quantity $\varrho= \frac{3}{2}\frac{k+\alpha-\lambda}{k+\alpha}$, and demanding that $\phi_{n}(x^\mu)$ satisfies the four-dimensional massive Klein--Gordon equation
$(\partial_{\mu} \partial^{\mu} -m^2_{n})\phi_{n}=0$,
we obtain the equation of the extra-dimensional part $\chi_n (z)$ of the scalar KK mode, which can be converted into  the following Schr\"{o}dinger-like equation:
\begin{equation}
\left[ -\partial_z^2 + V_{\text{s}}(z) \right] \chi_n(z)= m_n^2 \chi_n(z), \label{STbrane_scalar_equation}
\end{equation}
where $m_n$ is the mass of the scalar KK mode $\phi_n(x^\mu)$ and the effective potential $V_{\text{s}}(z)$ reads
\begin{eqnarray}
 V_{\text{s}}(z)=\varrho\partial_{z}^2 A(z)
        +[\varrho\partial_{z}A(z)]^2.
\label{STbrane_KK_scalar_potential}
\end{eqnarray}

It is clear that the effective potential $V_s$ is only dependent on the warp factor. In fact, the
Schr\"{o}dinger-like equation~(\ref{STbrane_scalar_equation}) can be written as
$\mathcal{H}\chi_{n}= m_{n}^{2}\chi_{n}$, where the Hamiltonian operator is given by
$\mathcal{H}=Q^{+}Q$ with $Q=-\partial_{z}+\varrho\partial_{z}A$. Since the operator
$\mathcal{H}$ is positive definite, there are no KK modes with negative $m_{n}^{2}$.

By substituting the KK decomposition \eqref{kk decompotition} into the five-dimensional scalar action  (\ref{STbrane_scalar_action}) and introducing the following orthonormality conditions:
\begin{eqnarray}
\int_{-z_{b}}^{+z_{b}} \chi_m(z) \chi_n(z)dz=\delta_{mn}, \label{STbrane_scalar_orthonormality_condition 1}
\end{eqnarray}
we get the effective low-energy theory in a four-dimensional familiar form
\begin{eqnarray}
 S_{\text{s}} = -\frac{1}{2}\sum_n \int d^4 x ( \eta^{\mu\nu} \partial_\mu \phi_n \partial_\nu \phi_n + m_n^2 \phi_{n}^2 ), \label{STbrane_scalar_four-dimensional_action}
\end{eqnarray}
where we have used Eq. (\ref{STbrane_scalar_equation}).

Due to the explicit forms of $\varrho$ and $A(z)$, the expression of the effective potential $V_{\text{s}}(z)$ corresponding to Eq. (\ref{STbrane_KK_scalar_potential}) is given by
\begin{eqnarray}
 V_{\text{s}}(z)&=&\frac{[(k-\lambda)^2-\alpha^2]\beta^2}{4 \alpha^2 (1+ \beta |z|)^2}\nonumber\\
                              &&+s\beta \Big[\delta(z)-\frac{1}{1+\beta z_b}\delta(z-z_b)\Big],
           \label{STbrane_scalar_potential_V0}
\end{eqnarray}
where $s=\frac{\alpha+k-\lambda}{\alpha}$. The value of $V_{\text{s}}(z)$ at $z = 0$ is
\begin{eqnarray}
V_{\text{s}}(0)&=&\frac{[(k-\lambda)^2-\alpha^2]\beta^2}{4 \alpha^2}
        +s\beta\delta(0).
\end{eqnarray}
In order to localize the scalar zero mode  on the positive tension brane, the effective potential $V_{\text{s}}(z)$ should be negative at $z=0$, which results in the following condition:
\begin{eqnarray}
 \lambda >k+\alpha.
    \label{STbrane_scalar_localized_condition}
\end{eqnarray}
By setting $m=0$ in Eq. (\ref{STbrane_scalar_equation}), we easily get the normalized zero mode of the scalar field for $\lambda\neq k+2\alpha$:
\begin{eqnarray}
 \chi_{0}(z)= \sqrt {\frac{\beta (k+2\alpha-\lambda)}
                           {2\alpha\left[(1+\beta z_b)^\frac{(k+2\alpha-\lambda)}{\alpha}-1\right]}}
                     \left(1+\beta |z| \right)^\frac{k+\alpha-\lambda}{2\alpha},
             \label{STbrane_scalar_zero_mode}
\end{eqnarray}
which is indeed localized  on the positive tension brane under the condition (\ref{STbrane_scalar_localized_condition}). When the extra dimension is infinite, the normalization condition $\int^{+\infty}_{-\infty}\chi_0(z) ^{2}dz=1$ means that the localization condition for the scalar zero mode is much stronger:
$\lambda> k+2\alpha$ .
For $\lambda= k+2\alpha$ , the normalized scalar zero mode reads as
\begin{eqnarray}
   \chi_{0}(z)=\sqrt{\frac{\beta}{2 \ln(1+\beta z_b)}}\frac{1}{\sqrt{1+\beta |z|}}~~~
   (~\lambda= k+2\alpha),
 \label{STbrane_scalar_special_zero_mode}
\end{eqnarray}
and it is also localized  on the positive tension brane. But if the extra dimension is infinite, the scalar zero mode is not normalized and hence we cannot get a localized scalar zero mode. When there is no coupling between the scalar and dilaton fields, i.e., the parameter $\lambda=0$, one can verify that the scalar zero mode is also localized on the positive tension brane when $z_b$ is finite.

In order to investigate the mass spectrum of the scalar KK modes, we redefine the following dimensionless parameters:
\begin{equation}
\bar{|z|}\equiv\beta|z|,~\bar{m}_{n}\equiv\frac{m_{n}}{\beta},~
\bar{V}_{s}(\bar{z})\equiv\frac{V_{s}(z)}{\beta^{2}},
\label{scalar dimensionless redefine}
\end{equation}
with which the Schr$\ddot{o}$dinger-like equation (\ref{STbrane_scalar_equation}) is rewritten as
\begin{equation}
 (-\partial^{2}_{\bar{z}}+\bar{V}_{s})\chi(\bar{z})=\bar{m}_{n}^{2}\chi(\bar{z}),
 \label{STbrane_scalar equation11}
\end{equation}
where the effective potential $\bar{V}_{s}$ reads as
\begin{equation}
 \bar{V}_{\text{s}}(z)=\frac{[(k-\lambda)^2-\alpha^2]}{4 \alpha^2 (1+ |\bar{z}|)^2}
        + c
          \Big[\delta(\bar{z})-\frac{1}{1+\bar{z}_b}\delta(\bar{z}-\bar{z}_b)\Big].
           \label{STbrane_scalar_effecitve potential_VS}
\end{equation}

{The general solution of Eq. \eqref{STbrane_scalar equation11} is given in terms of the combination of Bessel functions
\begin{eqnarray}
 \chi_n(\bar{z})&=&N_{n}(1+|\bar{z}|)^\frac{1}{2}
                         \Big[ C_{1}\text{J}_{P_{\text{s}}}(\bar{m}_{n}(|\bar{z}|+1))\nonumber\\
         &&~~~~+C_{2}\text {Y}_{P_{\text{s}}}(\bar{m}_{n}(|\bar{z}|+1))
     \Big],
 \label{STbrane_scalar_massive_Solution}
\end{eqnarray}
where $N_{n}$ is the normalization coefficient, $C_{1}$ and $C_{2}$ are the $m$-dependent parameters, $\text{J}_{P}(z)$ and $\text{Y}_{P}(z)$ are the Bessel functions of the first and second kinds.

Then we impose the two kinds of boundary condition (the Neumann boundary condition and the Dirichlet boundary condition) to calculate the mass spectra of scalar KK modes, respectively. For the the Neumann boundary condition $\partial_{\bar{z}}(e^{-\varrho A(\bar{z})} \chi_n)|_{z=0,z=z_{b}}=0$, we get the mass spectrum which is determined by following condition:
\begin{eqnarray}
 \mathcal{M}_{N}(\bar{m}_{n})\equiv &&\text{J}_{P_{\text{Ns}}} (\bar{m}_{n}(\bar{z}_b+1))\nonumber\\
               &&~~~~~+\mathcal{C}_{\text{Ns}}\text{Y}_{P_{\text{Ns}}} (\bar{m}_{n}(\bar{z}_b+1))=0 .
               \label{STsmcondition1}
\end{eqnarray}
where $P_{\text{Ns}}$ and $  \mathcal{C}_{\text{Ns}}$ are defined as:
\begin{eqnarray}
  {P_{\text{Ns}}}\equiv {\frac{\lambda-k-2\alpha}{2\alpha}},~~~
  \mathcal{C}_{\text{Ns}}\equiv
           -\frac{\text{J}_{P_{\text{Ns}}}(\bar{m}_n)}
                    {\text{Y}_{P_{\text{Ns}}}(\bar{m}_n)}. \label{PsCs1}
   \end{eqnarray}
And, for the Dirichlet boundary condition $e^{-\varrho A(\bar{z})} \chi_n=0$ at the boundaries $\bar{z}=0$ and $\bar{z}=\bar{z}_{b}$, we get  the mass spectrum of the scalar KK modes which satisfies the following relation:
\begin{eqnarray}
 \mathcal{M}_{D}(\bar{m}_{n})\equiv &&\text{J}_{P_{\text{Ds}}} (\bar{m}_{n}(\bar{z}_b+1))\nonumber\\
               &&~~~~~+\mathcal{C}_{\text{Ds}}\text{Y}_{P_{\text{Ds}}} (\bar{m}_{n}(\bar{z}_b+1))=0 .
               \label{STsmcondition2}
\end{eqnarray}
where $P_{\text{Ds}}$ and $\mathcal{C}_{\text{Ds}}$ are given by 
\begin{eqnarray}
  {P_{\text{Ds}}}\equiv {\frac{\lambda-k}{2\alpha}},~~~
  \mathcal{C}_{\text{Ds}}\equiv
           -\frac{\text{J}_{P_{\text{Ds}}}(\bar{m}_n)}
                    {\text{Y}_{P_{\text{Ds}}}(\bar{m}_n)}. \label{PsCs2}
   \end{eqnarray}
From Eqs. \eqref{STsmcondition1} and \eqref{STsmcondition2},  we  see  that  the  Neumann and Dirichlet  boundary conditions give the similar expressions which  determine the mass spectrum, whereas the orders of the Bessel functions are different as shown in \eqref{STsmcondition1} and \eqref{PsCs1}. By tuning the parameters $k$ and $\lambda$, the same orders of Bessel functions can be obtained, and thus the expression of the mass spectrum derived by the
Neumann boundary condition is equivalent to that derived by the Dirichlet boundary condition. Therefore, we
only consider the Neumann boundary condition for later discussion.

According to Eq. \eqref{STsmcondition1}, we plot the relation between $\mathcal{M}_{N}$ and $\bar{m}_{n}$ in Figs. \ref{STbrane_fig_Spectrum_Scalar1}, \ref{STbrane_fig_Spectrum_Scalar2}, \ref{STbrane_fig_Spectrum_Scalar3}, where the zero points represent the mass of the KK modes.
Figure \ref{STbrane_fig_Spectrum_Scalar1} shows that the effect of the parameter $\lambda$ on the mass $\bar{m}_1$ of the first massive scalar KK mode. It can be seen that the mass $\bar{m}_{1}$ increases with $\lambda$. On the other hand, there are some singular (vertical) lines in these figures, which come from the zero points of ${\text{Y}_{P_{\text{Ns}}}(\bar{m}_n)}$ in \eqref{PsCs1}. Since the gaps between two arbitrary adjacent singular lines are almost the same for a set of parameters, we can call the average gap the period $T$. For example, the periods $T$ for the three cases shown in Fig. \ref{STbrane_fig_Spectrum_Scalar1} are 3.13, 3.22 and 3.42, respectively. The period for a special mass spectrum increases with $\lambda$.
Figure \ref{STbrane_fig_Spectrum_Scalar2} states the effect of the size of the extra dimension, $\bar{z}_{b}$, on the gap of the massive KK modes. It shows that the number of excited states in a single period increases with $\bar{z}_{b}$, and the gap of massive KK modes decreases with $\bar{z}_{b}$. An interesting phenomenon will appear if we set $\bar{z}_{b}<1$ (and $k=-3$ and $\lambda=5$ at the same time). The massive KK modes will not appear in every period anymore for this case; however, they will appear after severe periods, as shown in Fig. \ref{STbrane_fig_Spectrum_Scalar3}. The mass spectra of the scalar KK modes are shown in Fig.~\ref{STbrane_fig_Spectrum_Scalar4} for different parameters. The mass gap between the zero mode and the first massive KK mode increases with the dilaton coupling constant $\lambda$ and decreases with the size of the extra dimension $\bar{z}_{b}$. On the other hand, the mass spectrum is almost equidistant for the higher excited states, and relatively sparse for the lower excited states.
\begin{figure}[htb]
\subfigure[$\lambda=1$]{
\includegraphics[width=0.31\textwidth]{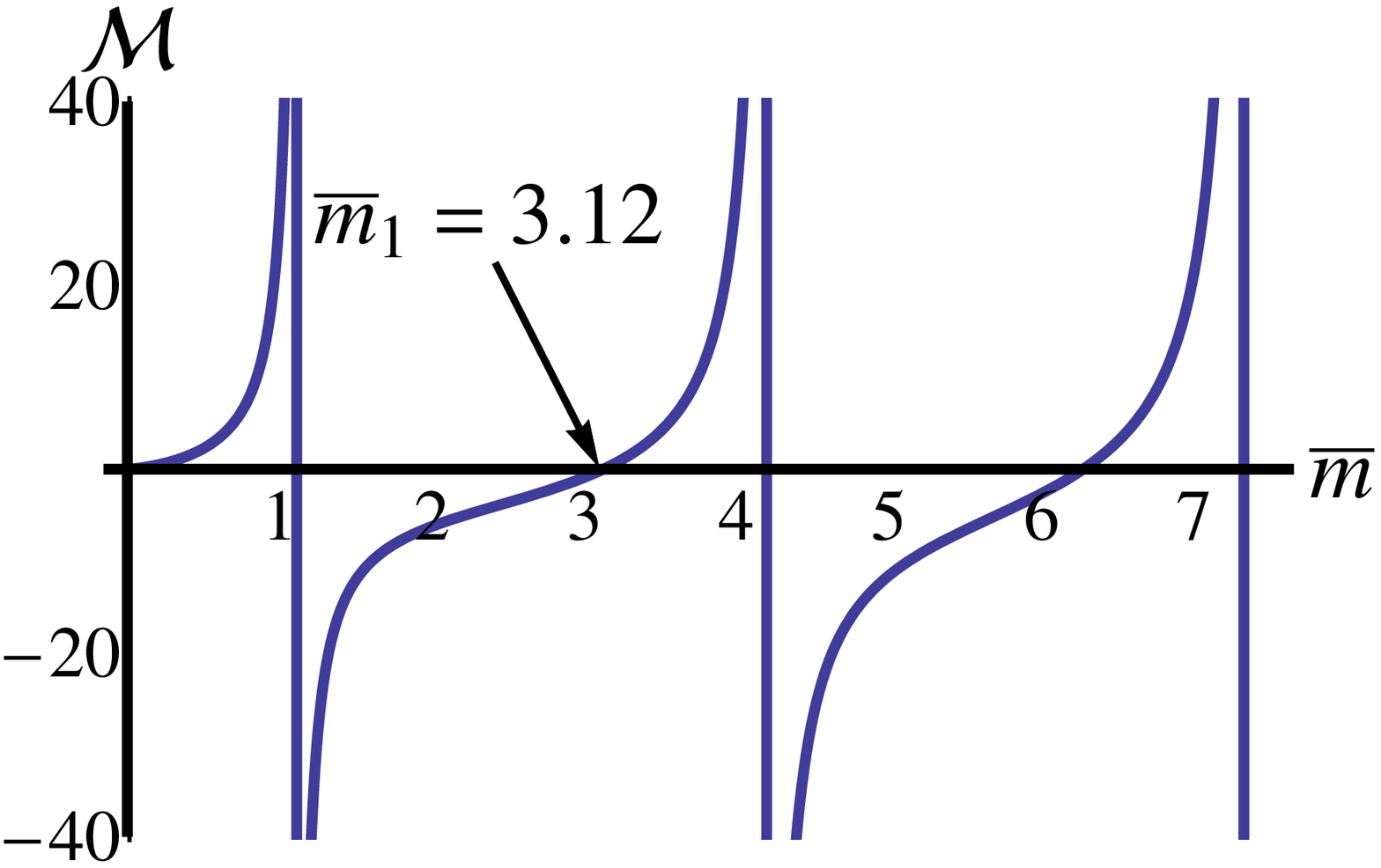}}
\subfigure[$\lambda=10$]{
\includegraphics[width=0.31\textwidth]{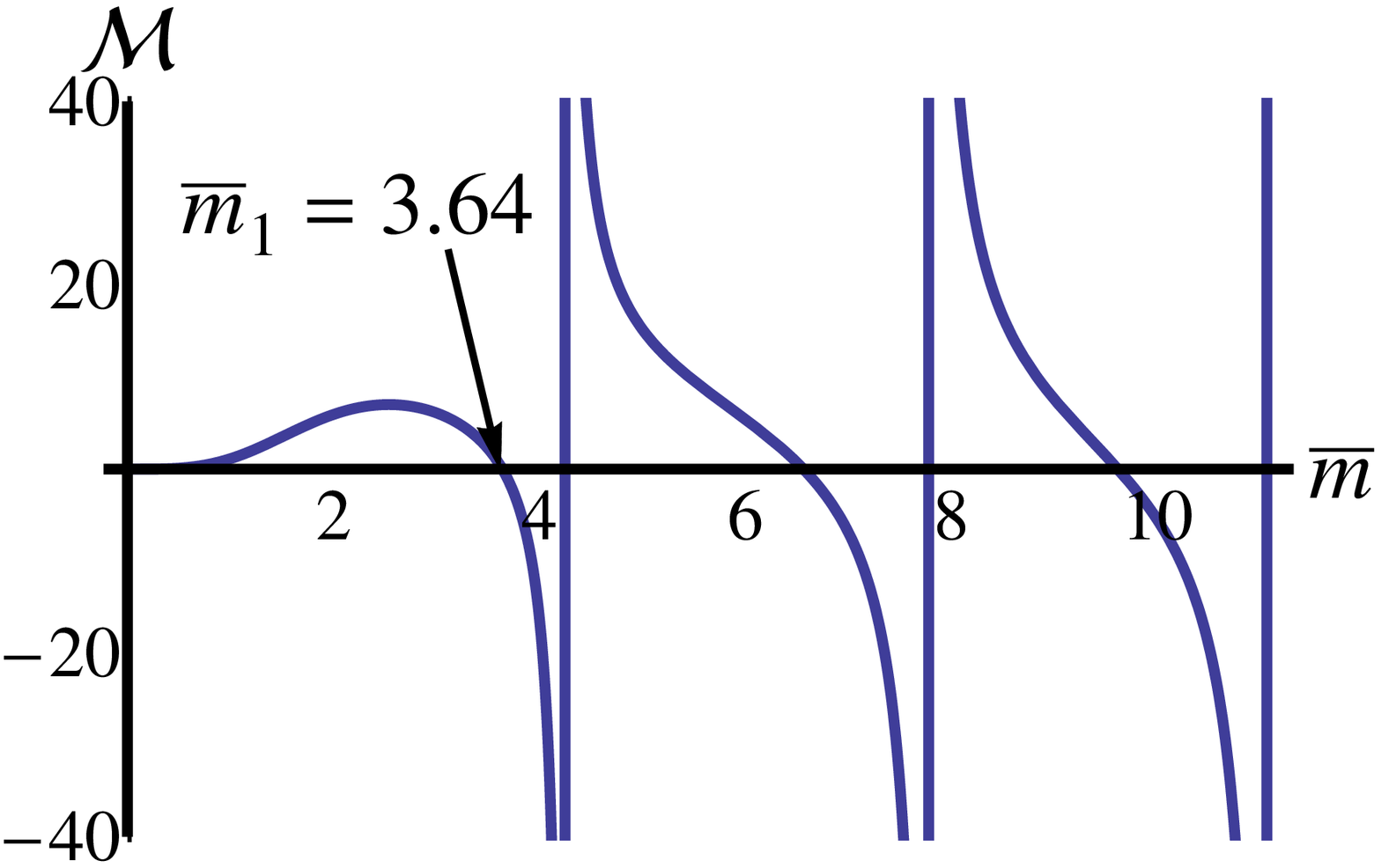}}
\subfigure[$\lambda=30$]{
\includegraphics[width=0.31\textwidth]{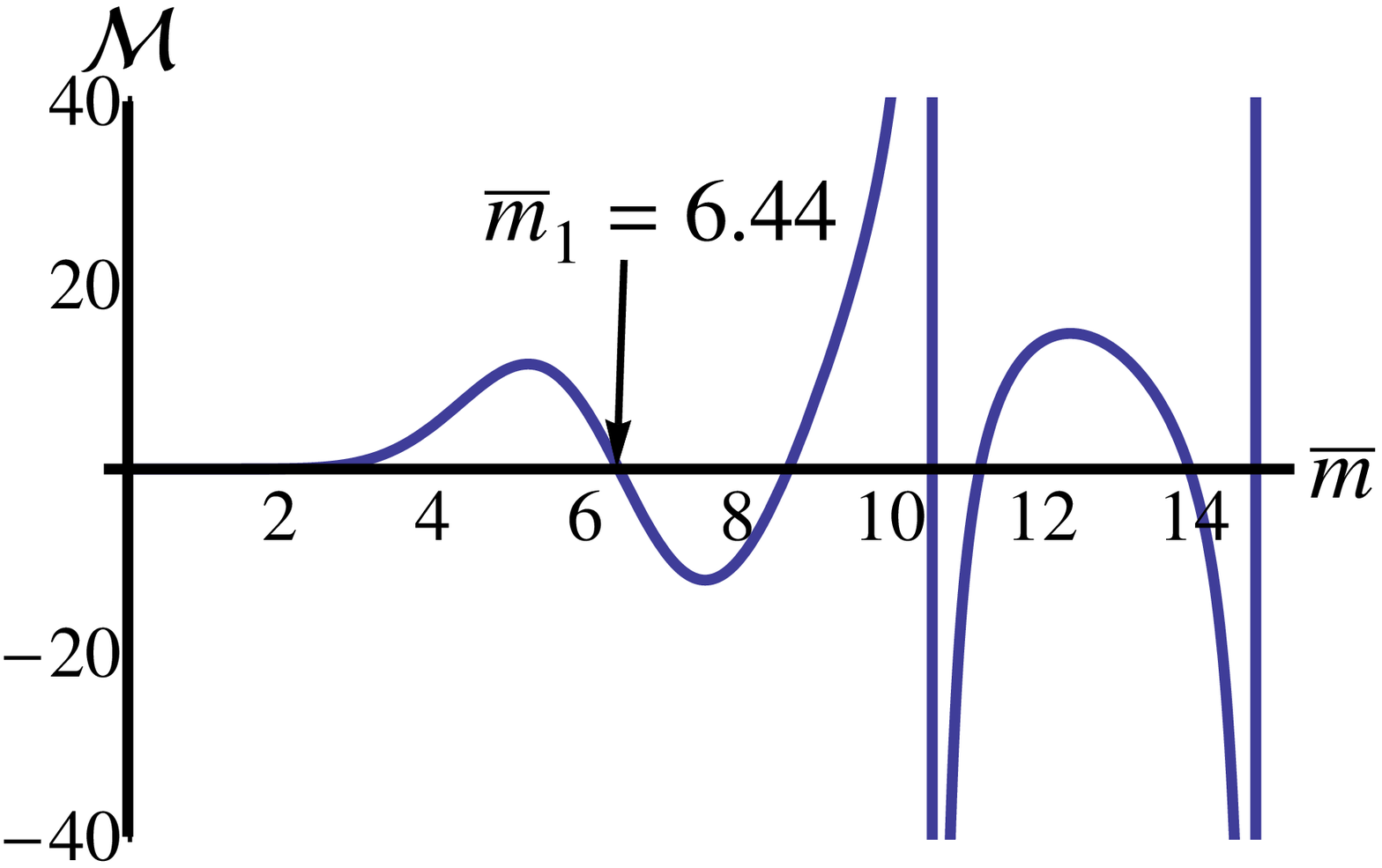}}
 \caption{The effect of the parameter $\lambda$ on the mass $\bar{m}_1$ of the first massive scalar KK mode.
 The other parameters  are set to $k=-3$ and $\bar{z}_b=1$.}
\label{STbrane_fig_Spectrum_Scalar1}
\end{figure}
\begin{figure}[htb]
\subfigure[$\bar{z}_{b}=2$]{
\includegraphics[width=0.31\textwidth]{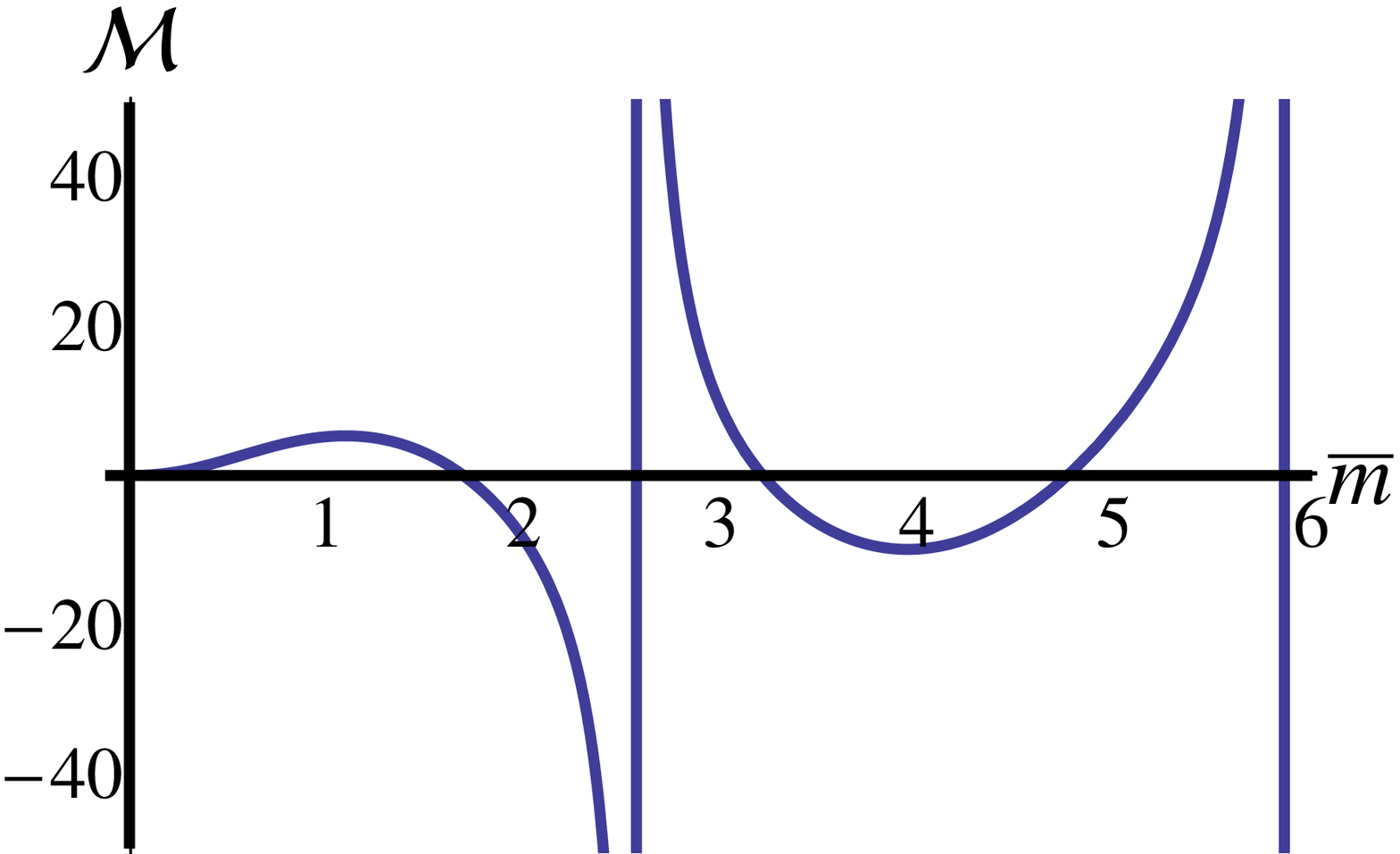}}
\subfigure[$\bar{z}_{b}=5$]{
\includegraphics[width=0.31\textwidth]{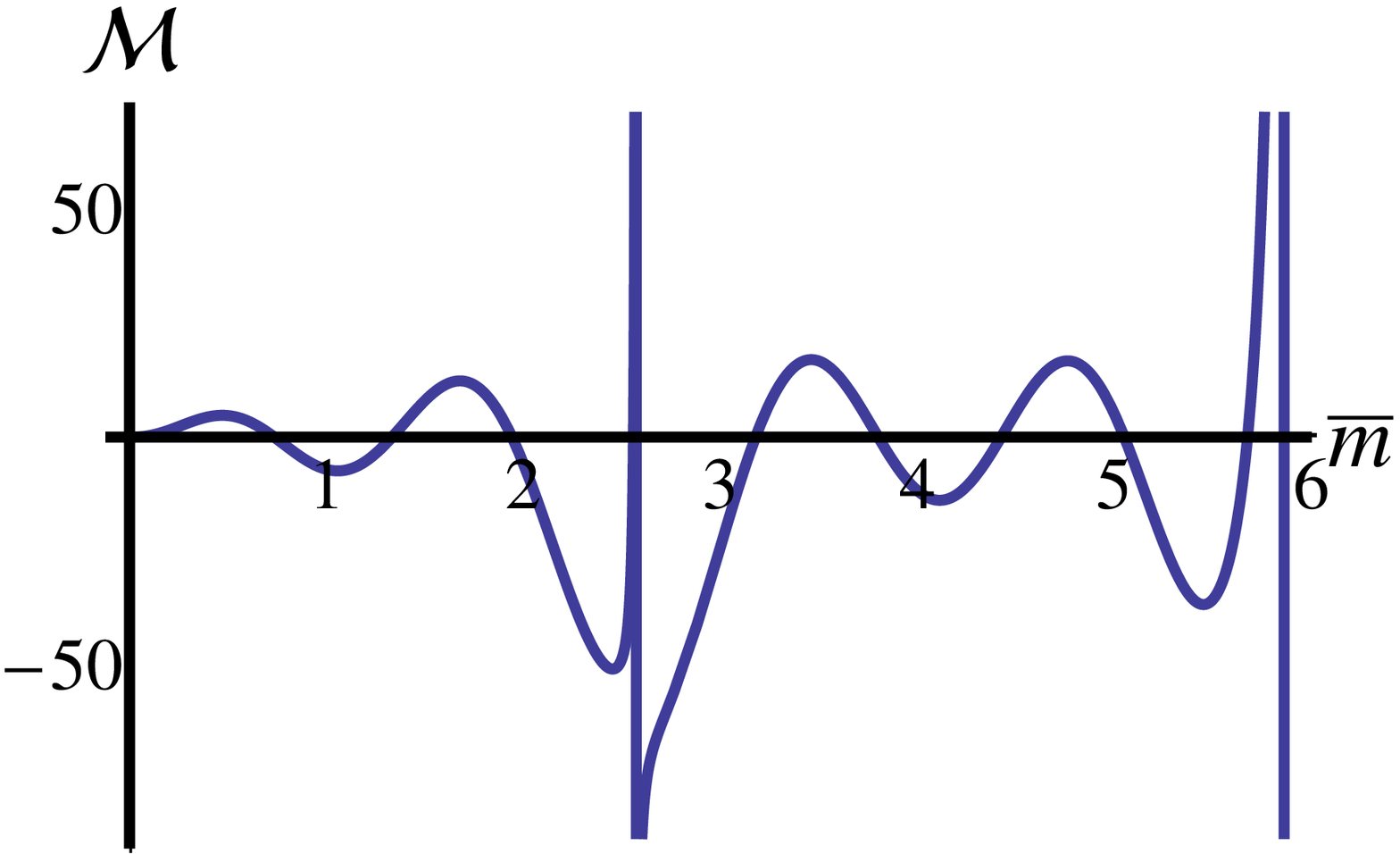}}
\subfigure[$\bar{z}_{b}=10$]{
\includegraphics[width=0.31\textwidth]{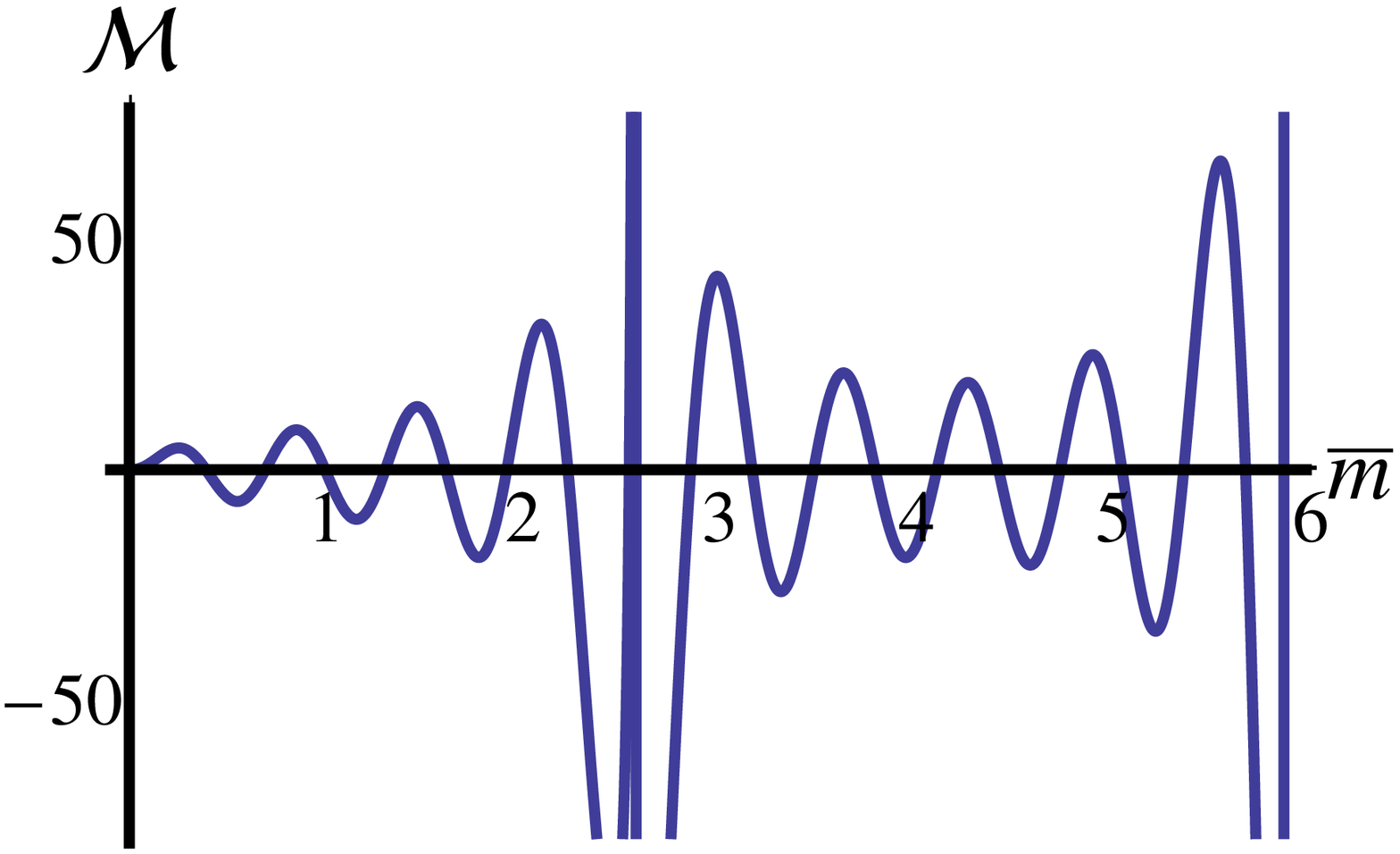}}
 \caption{The effect of the size of the extra dimension, $\bar{z}_b$ (large $\bar{z}_b>1$), on the mass spectra of the scalar KK modes.
 The other parameters  are set to $k=-3$ and $\lambda=5$.}
\label{STbrane_fig_Spectrum_Scalar2}
\end{figure}

\begin{figure}[htb]
\subfigure[$\bar{z}_{b}=0.1$]{
\includegraphics[width=0.43\textwidth]{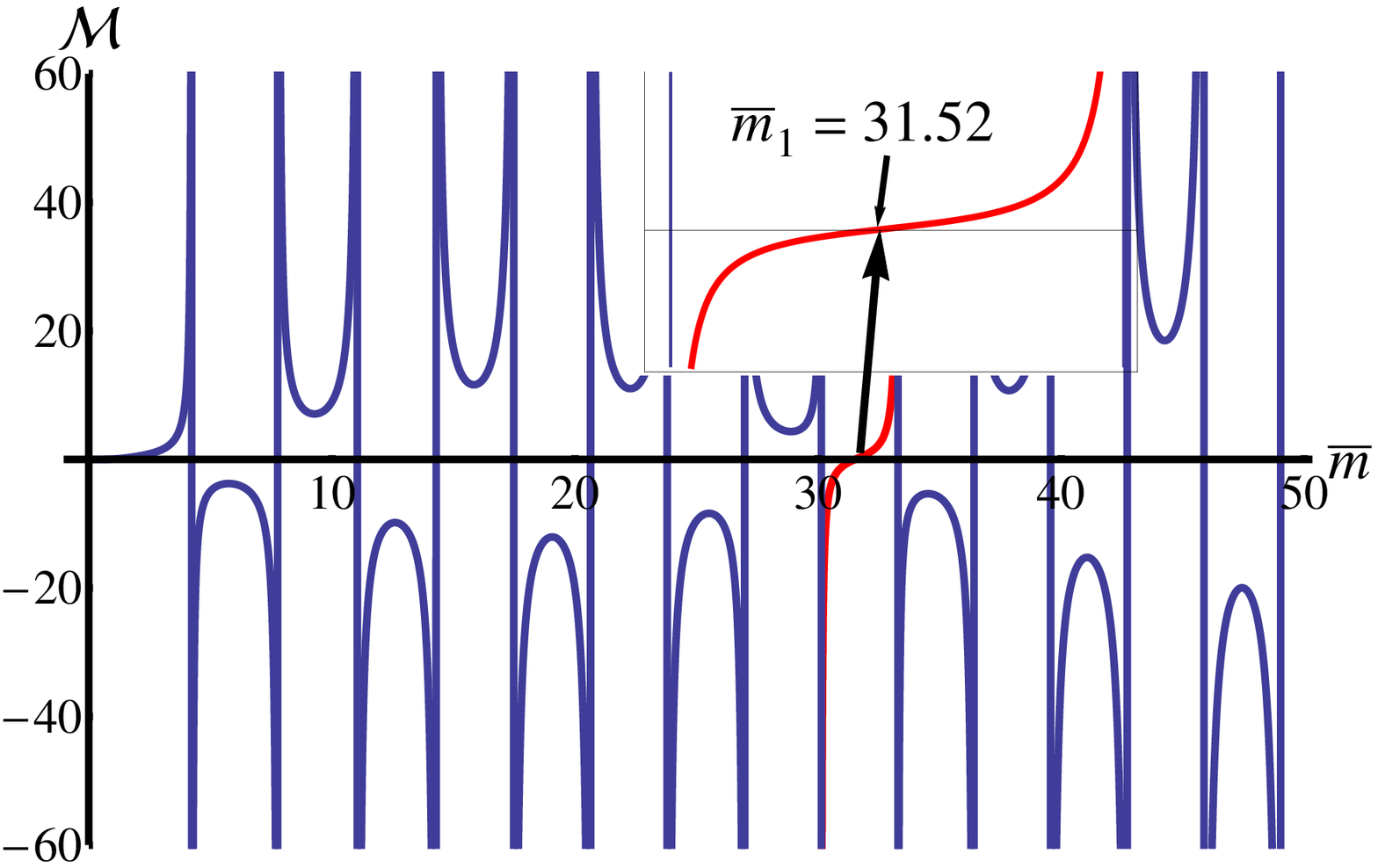}}
\subfigure[$\bar{z}_{b}=0.05$]{
\includegraphics[width=0.43\textwidth]{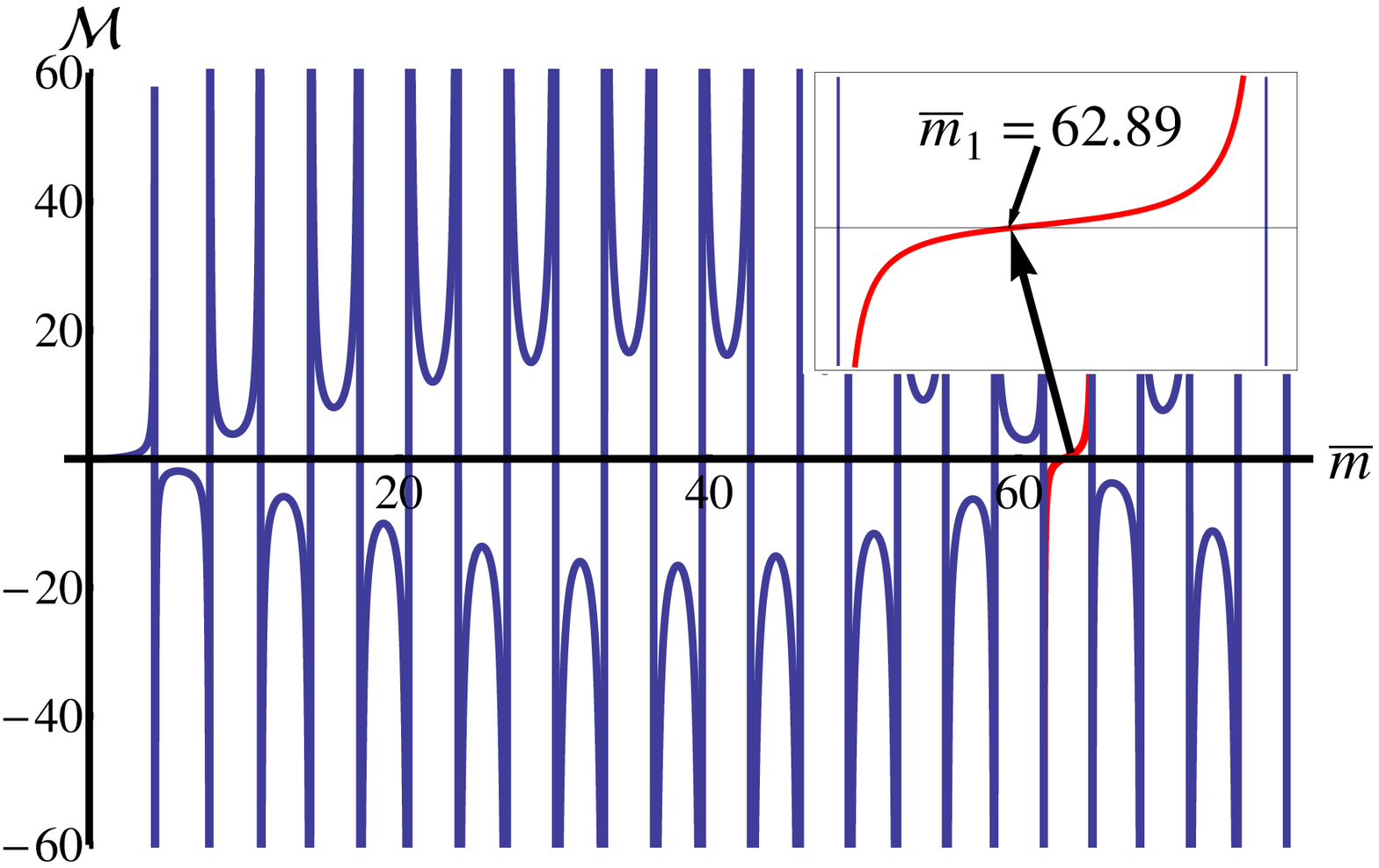}}
\caption{The effect of the size of the extra dimension, $\bar{z}_b$ (small $\bar{z}_b <1$), on the mass spectra of the scalar KK modes. The other parameters are set to $k=-3$, $\lambda=5$.}
\label{STbrane_fig_Spectrum_Scalar3}
\end{figure}

\begin{figure}[htb]
\subfigure[$\bar{z}_{b}=50$]{
\includegraphics[width=0.43\textwidth]{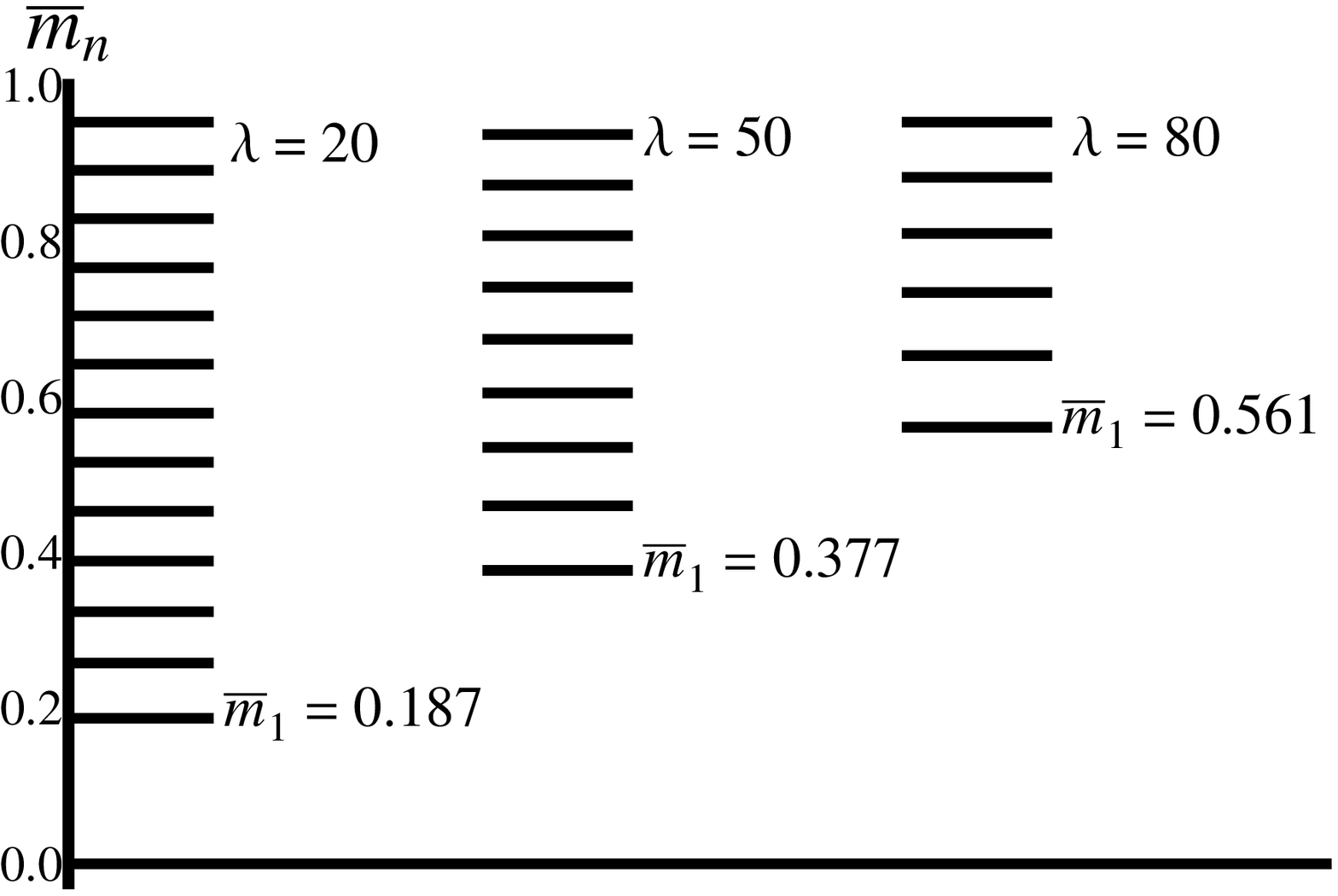}}
\subfigure[$\lambda=80$]{
\includegraphics[width=0.43\textwidth]{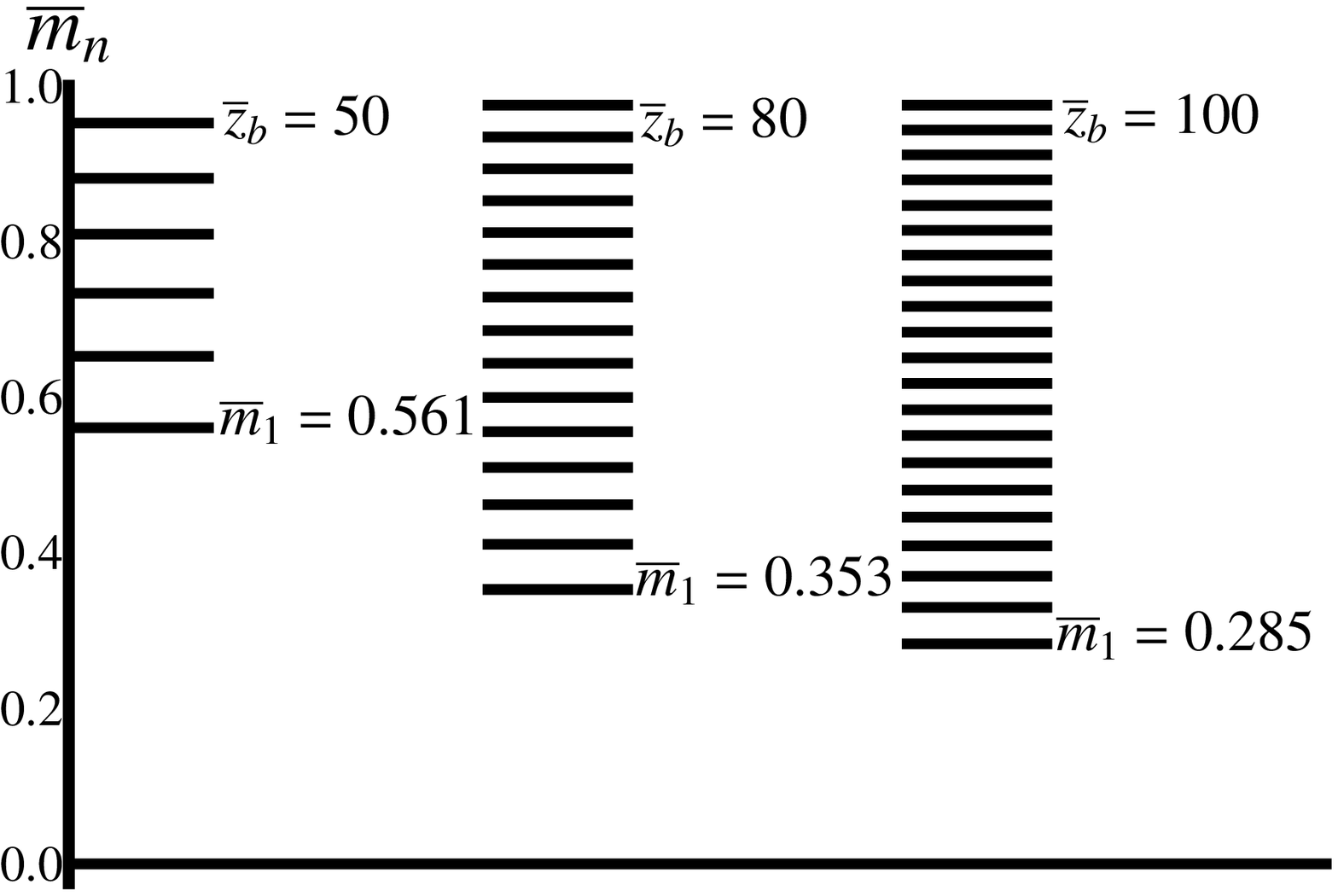}}
\caption{The mass spectra of the scalar KK modes. The parameter $k$ is set to $k=-3$.}
\label{STbrane_fig_Spectrum_Scalar4}
\end{figure}

\subsubsection{The Higgs potential coupling}\label{3.1.2}
Next, we consider another coupling, the Higgs potential type coupling between a real scalar $\Phi$ and the background scalar field $\omega$. The five-dimensional action is assumed as \cite{Liu2009uca},
\begin{equation}
S_{\text{s}} = \int d^5 x  \sqrt{-g}\Big[-\frac{1}{2} g^{MN} \partial_M \Phi \partial_N \Phi-V(\Phi,\omega)\Big],
\label{STbrane_scalar_action 2}
\end{equation}
where the Higgs potential $V(\Phi,\omega)$ is given by
\begin{equation}
V(\Phi,\omega)=(\theta\omega^2-u^2)\Phi^2.
\label{STbrane_scalar_potential}
\end{equation}
According to the metric (\ref{STbrane_linee1}) and the action (\ref{STbrane_scalar_action 2}), the equation of motion of the five-dimensional  real scalar $\Phi$ reads
\begin{equation}
\partial_{\mu}\partial^{\mu}\Phi
 +e^{-3A}\partial_{z}(e^{3A}\partial_{z}\Phi)
 -2 e^{2A} (\theta\omega^2-u^2) \Phi=0.
\label{eqnarray2}
\end{equation}
Then, making  the KK decomposition\\
$\Phi(x^\mu,z)=\sum_n\phi_{n}(x^\mu)\chi_{n}(z)e^{-3A/2}$
and demanding that $\phi_{n}(x^\mu)$ satisfies the four-dimensional massive Klein--Gordon equation $(\partial_{\mu} \partial^{\mu} -m^2_{n})\phi_{n}=0$,  we obtain the Schr$\ddot{o}$dinger-like equation of the extra--dimensional part $\chi_{n}(z)$
\begin{equation}
\left[ -\partial_z^2 + V_{\text{s}}(z) \right] \chi_n(z)= m_n^2 \chi_n(z). \label{STbrane_scalar_equation2}
\end{equation}
Here  $m_{n}$ is the mass of the $n$-th KK mode, and the effective potential $V_{s}(z)$ is given by
\begin{equation}
V_{s}(z)=\frac{3}{2}\partial^2_{z}A+\frac{9}{4}(\partial_{z}A)^2+2e^{2A}(\theta\omega^2-u^2).
\label{effective potential2}
\end{equation}
The fundamental five-dimensional action (\ref{STbrane_scalar_action 2}) can be reduced to the effective four-dimensional action of  the massive scalars, when $\chi_{n}$  satisfy the following orthonormality conditions:
\begin{eqnarray}
\int_{-z_{b}}^{+z_{b}}\chi_m(z) \chi_n(z)dz=\delta_{mn}. \label{STbrane_scalar_orthonormality_condition}
\end{eqnarray}
The explicit expression of the effective potential $V_{s}(z)$ for the brane solution \eqref{BraneSolution} is
\begin{eqnarray}
  V_{s}(z) &=&\frac{(k^2-\alpha^2)\beta^2}
                   {4\alpha^2(1+\beta|z|)^2}\nonumber\\
               &&+ 2(1+\beta|z|)^{\frac{2(\alpha+k)}{3\alpha}}
                 \Big(\frac{\theta}{\alpha^2}\ln(1+\beta|z|)-u^2\Big)  \nonumber\\
      &&+  \frac{k+\alpha}{\alpha}\beta
           \Big [\delta(z) -\frac{1}{1+\beta z_{b}}\delta(z-z_{b}) \Big].
\label{effective potential21}
\end{eqnarray}
At $z=0$, the value of $V_{s}(z)$ is
\begin{equation}
V_{s}(0)=\frac{(k^2-\alpha^2)\beta^2-8u^2\alpha^{2}}{4\alpha^2} +\frac{k+\alpha}{\alpha}\beta\delta(0).
\end{equation}
Considering that $\beta>0$, $k<-1$ and $\sqrt{3}/2<\alpha=\sqrt{k^2+3k+3}<-k$, we know that the effective potential $V_{s}(z)$ is negative at $z=0$. Therefore, the scalar zero mode can be localized on the positive tension brane.

For convenience, we redefine some new dimensionless parameters:
  \begin{equation}
  |\bar{z}|\equiv\beta |z|~,\bar{\theta}\equiv\frac{\theta}{\beta^2}~,\bar{u}\equiv\frac{u}{\beta}~,
  \bar{m}_{n}\equiv\frac{m_{n}}{\beta},
  \end{equation}
and the Schr$\ddot{o}$dinger-like equation (\ref{STbrane_scalar_equation2}) is changed as
\begin{equation}
  (-\partial^{2}_{\bar{z}}+\bar{V}_{s})\chi(\bar{z})=\bar{m}_{n}^{2}\chi(\bar{z}),
 \label{STbrane_scalar_equation3}
\end{equation}
where the expression of the dimensionless effective potential $\bar{V}_{s}$ is
\begin{eqnarray}
\bar{V}_{s}(\bar{z})\equiv\frac{V_s}{\beta^{2}}
   &=&\frac{(k^2-\alpha^2)}{4\alpha^2(1+|\bar{z}|)^2}\nonumber\\
      &&+  2(1+|\bar{z}|)^{\frac{2(\alpha+k)}{3\alpha}}
         \left( \frac{\bar{\theta}}{\alpha^2}\ln(1+|\bar{z}|)-\bar{u}^2
         \right)\nonumber\\
   &&+ \frac{k+\alpha}{\alpha}
      \left[\delta(\bar{z})
            -\frac{1}{1+\bar{z}_{b}}\delta(\bar{z}-\bar{z}_{b})
      \right].
   \label{effective potentialVs3}
\end{eqnarray}
At the boundary $\bar{z}=\bar{z}_{b}$, the effective potential $\bar{V}_{s}(\bar{z})$ (\ref{effective potentialVs3}) has a positive $\delta$ function (just as an infinite high barrier), and it will yield infinite discrete bound KK modes. Due to the non-analytical solution of the Schr$\ddot{o}$dinger-like equation (\ref{STbrane_scalar_equation3}), we can only solve numerically Eq. (\ref{STbrane_scalar_equation3}) by fixing the parameter $\bar{\theta}$ and choosing the proper value $\bar{u}$, to ensure the zero mode can be localized on the brane.

The wave functions of the lower bound KK modes are plotted in Fig.~\ref{STbrane_fig_Spectrum_Scalar_Solution2} and the corresponding mass spectrum is listed as follows:
\begin{eqnarray}
\bar{m}_{n}&=&\big({0, ~2.84,~ 4.50, ~6.09, ~7.68, }\nonumber\\
      &&~9.25, ~10.81, ~12.36, ~13.91, ~15.46,\cdots\big),
\label{spectrum of the KK modes}
\end{eqnarray}
where the parameters are set to $k= -2,~\bar{\theta}=4.77,~\bar{z}_{b}=1$, and $\bar{u} = 1.43$. From the numerical solution above, we know that the zero mode can be  trapped on the brane by fine tuning the parameters. We can see that the mass spectrum would get denser for the higher excited states while they get sparser for the lower excited states, and the gap of the higher excited states is almost the same for a set of parameters, on account of the dimensionless effective potential $\bar{V}_{s}(\bar{z})$ at the  boundary $\bar{z}=\pm\bar{z}_{b}$ having two positive $\delta$ functions, and the effective potential $\bar{V}_{s}(\bar{z})$ just like an infinite high barrier. Therefor, the solutions at the high excited states should be trigonometric functions with the same mass gap.
By increasing the value of the parameter $\bar{u}$,  the bound state solution with $\bar{m}^{2}_{n}<0$ will appear. Therefore, in order to obtain nonnegative eigenvalues $\bar{m}^{2}$,  the mass parameter $\bar{u}$ will have an upper limit when the other parameters are fixed. For $k=-2,~\bar{\theta}=4.77,~\bar{z}_b=1$, as we have chosen here, the upper limit of $\bar{u}$ is $1.43$.
\begin{figure}[htb]
\subfigure[$\bar{m}_0=0$]{
\includegraphics[width=0.23\textwidth]{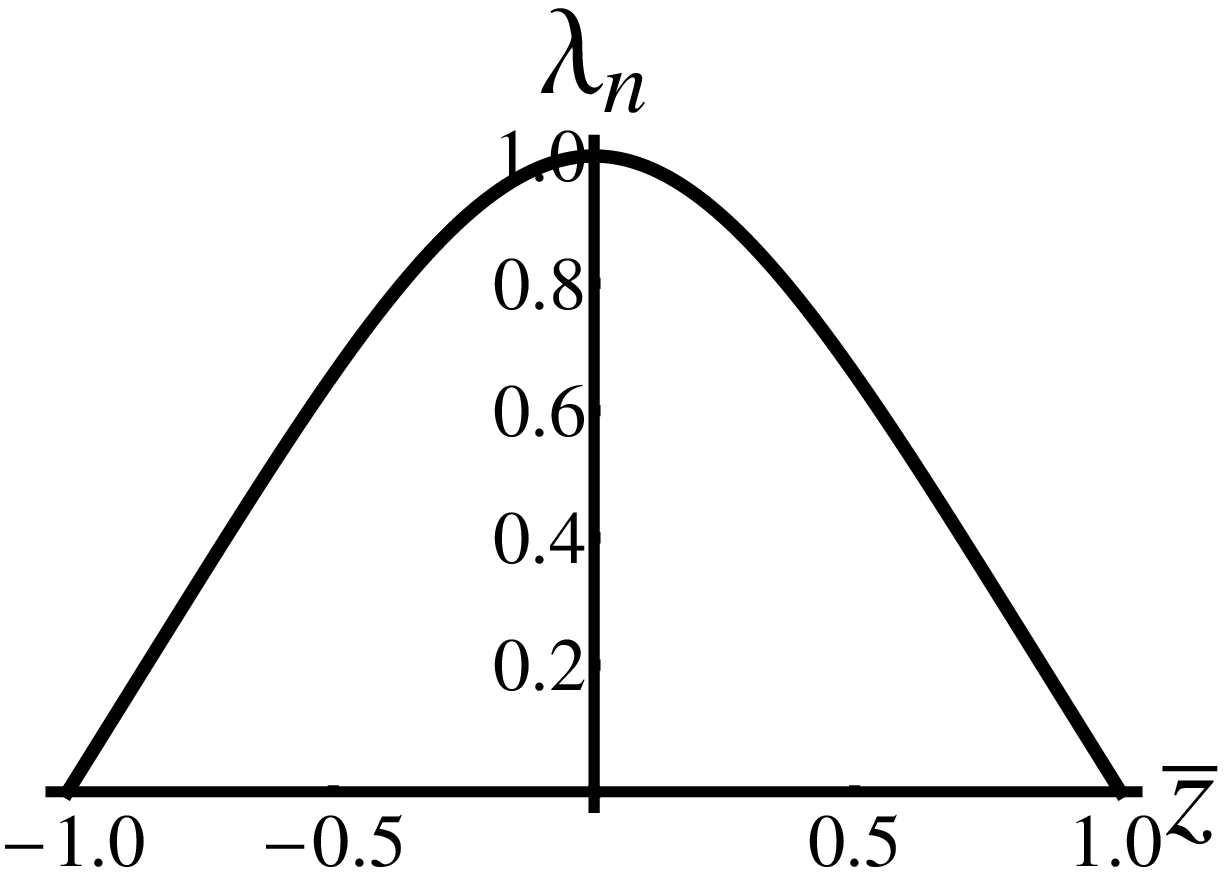}}
\subfigure[$\bar{m}_1=2.84$]{
\includegraphics[width=0.23\textwidth]{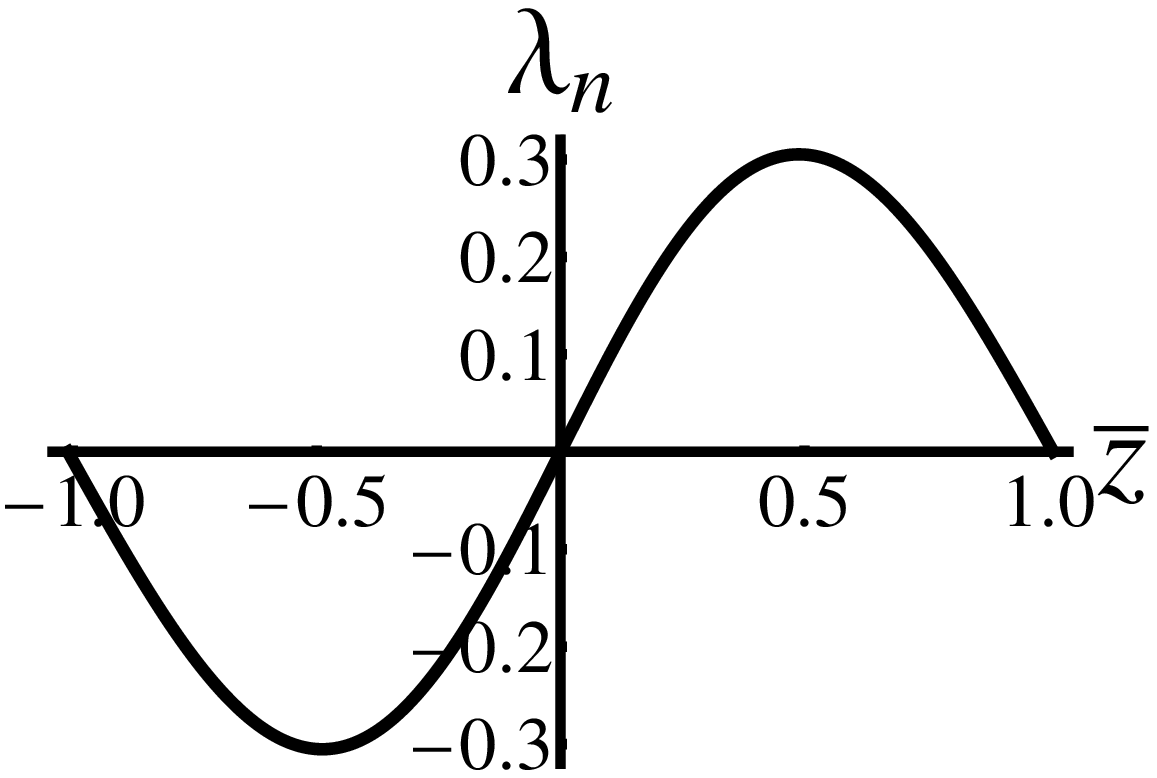}}
\subfigure[$\bar{m}_2=4.50$]{
\includegraphics[width=0.23\textwidth]{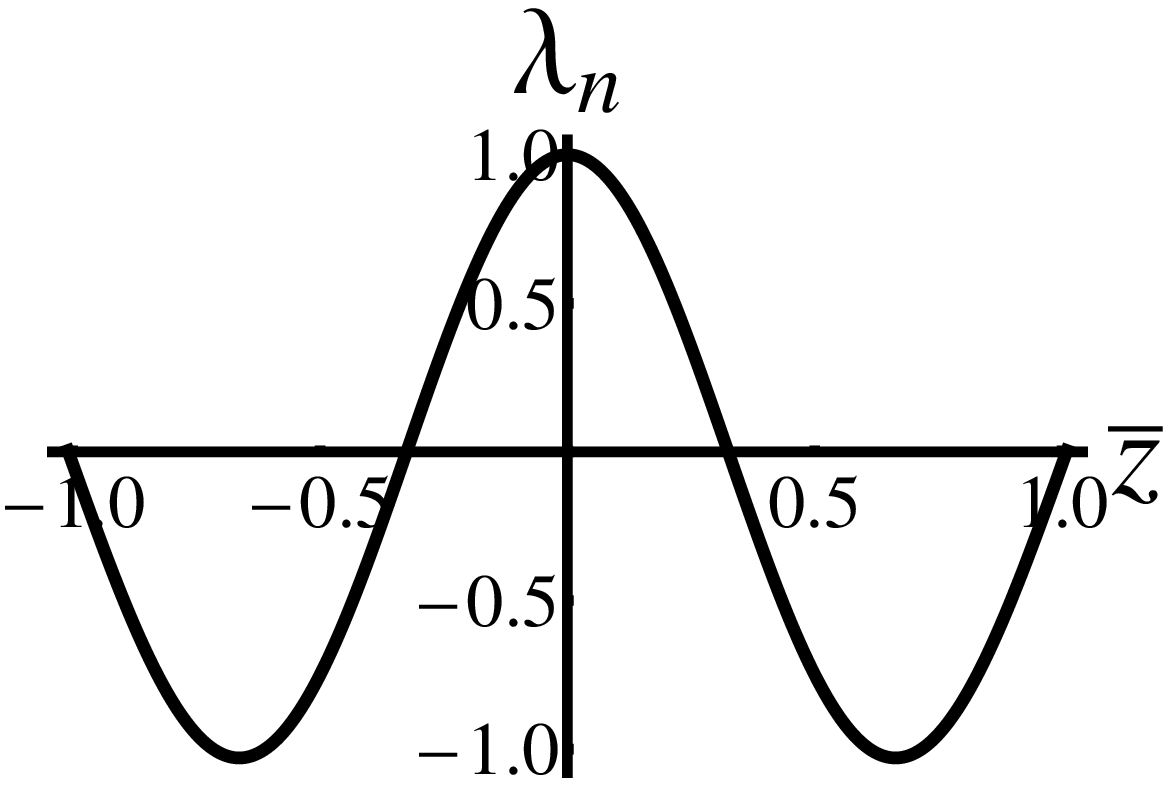}}
\subfigure[$\bar{m}_3=6.09$]{
\includegraphics[width=0.23\textwidth]{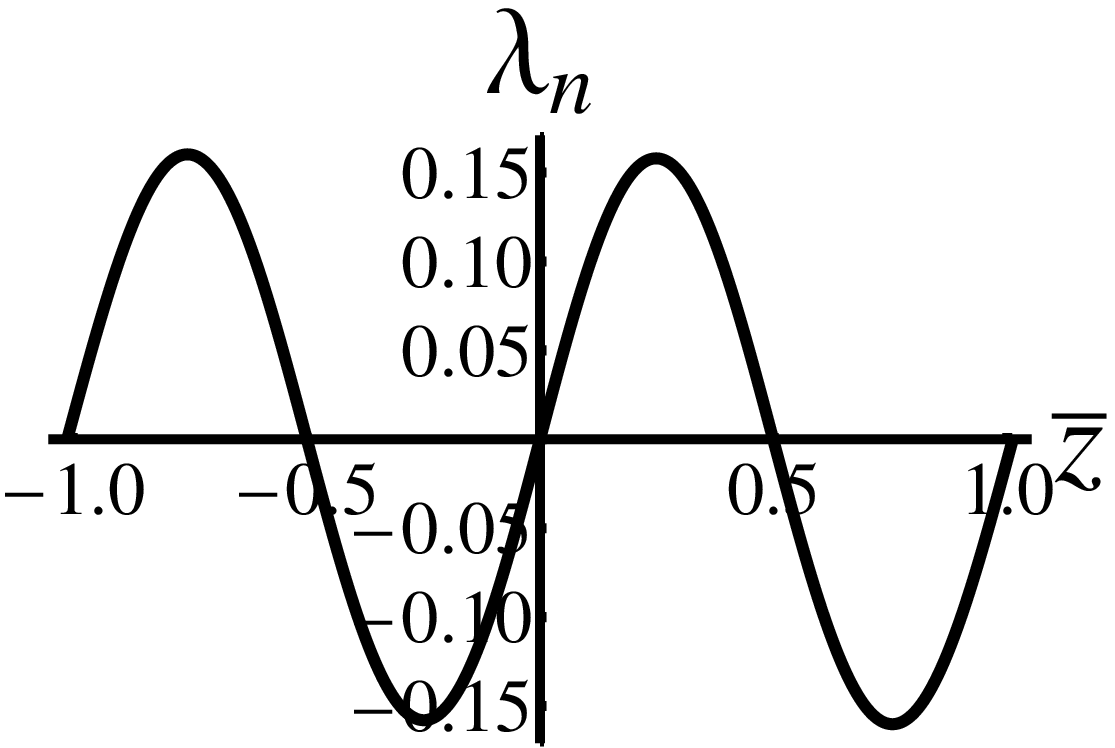}}
 \caption{The wave functions of the lower scalar KK modes.
 The parameters  are set to $k= -2,~\bar{\theta}=4.77,~\bar{z}_{b}=1$, and $\bar{u} = 1.43$.}
\label{STbrane_fig_Spectrum_Scalar_Solution2}
\end{figure}

\subsection{Spin-1 vector field}
\label{STbrane_secFermionZeroMode}

Secondly, we turn to the spin-$1$ vector field.    A typical  five-dimensional action of a
vector field coupled with a dilaton field $\omega$ is given as
\begin{equation}
S_{\text{v}} = -\frac{1}{4} \int d^5 x  \sqrt{-g} e^{\tau\omega}g^{MN} g^{RS}F_{MR}F_{NS},
\label{STbrane_vector_action1}
\end{equation}
where $F_{MN}=\partial_{M}A_{N}-\partial_{N}A_{M}$
is the field strength as usual and $\tau$ is a
dimensionless coupling constant. Considering the explicit form of the metric (\ref{STbrane_linee1}), the equations of motion read as
\begin{eqnarray}
  e^{\tau\omega}\partial_{\nu} F^{\mu\nu}
   +\eta^{\mu\lambda} a^{-1}(z)\partial_4 \left(a(z)e^{\tau\omega}F_{4\lambda}\right) &=& 0,   \nonumber \\
  e^{\tau\omega} \partial_\mu \left( \eta^{\mu\nu} F_{\nu4} \right)  &=& 0.
\end{eqnarray}

{In order to be consistent with the gauge invariant equation $\oint dz A_{4}=0$, we use the gauge freedom to choose $A_4=0$.}
Under the KK decomposition of the vector field
\begin{eqnarray}
 A_{\mu}(x^\mu,z)=\sum_n a^{(n)}_{\mu}(x^\mu) \rho_n(z)e^{-\epsilon A(z)},\label{fenjie}
\end{eqnarray}
where $\epsilon=\frac{k+\alpha-3\tau}{2(k+\alpha)}$, and the orthonormality conditions
\begin{eqnarray}
\int_{-z_{b}}^{+z_{b}}  \rho_m(z) \rho_n(z)dz=\delta_{mn},
\label{STbrane_vector_orthonormality_condition}
\end{eqnarray}
the action (\ref{STbrane_vector_action1}) can be reduced to an effective one including the four-dimensional massless vector field (the zero mode) and a set of massive vector fields (the massive KK modes):
 \begin{eqnarray}
 S_{\text{v}}=\sum_{n}\int d^4 x \Big(-\frac{1}{4}f^{(n)}_{\mu\nu}
 f^{{\mu\nu}(n)}
              -\frac{1}{2}m^2_{n}  a^{(n)}_{\mu} a^{{\mu}(n)}\Big),
 \label{STbrane_vector_four-dimensional_action}
\end{eqnarray}
where $f^{(n)}_{\mu\nu}=\partial_\mu a^{(n)}_\nu - \partial_\nu a^{(n)}_\mu$ is the field strength of the four-dimensional vector field. In addition, the extra--dimensional part $\rho_n (z)$ satisfies the following Schr\"{o}dinger-like equation:
\begin{equation}
\left[ -\partial_z^2 + V_{\text{v}}(z) \right] \rho_n(z)= m_n^2 \rho_n(z), \label{STbrane_vector_equation}
\end{equation}
where $m_n$ is the mass of the vector KK mode
and the effective potential $V_{\text{v}}(z)$  is
\begin{eqnarray}
 V_{\text{v}}(z)=(\epsilon \partial_{z}A)^2
        +\epsilon\partial_{z}^2 A.
 \label{STbrane_KK_vector_potential}
\end{eqnarray}
The explicit expression of the effective potential $V_{\text{v}}(z)$ reads
\begin{eqnarray}
 V_{\text{v}}(z)&=&\frac{b(k+\alpha-3\tau) \beta^2}{36\alpha^2 ( 1+ \beta |z|)^2}\nonumber\\
       && +\frac{(k+\alpha-3\tau)\beta }{3\alpha}\Big[ \delta(z)- \frac{\delta(z-z_b)}{1+\beta z_b} \Big],
 \label{STbrane_vector_potential_V1_1}
\end{eqnarray}
where b=$\left( k-5\alpha-3\tau\right)$.
The value of the effective potential $V_{\text{v}}(z)$ at $z = 0 $ is
\begin{eqnarray}
V_{\text{v}}(0)&=&\frac{b(k+\alpha-3\tau) \beta^2}{36 \left( k+\alpha \right)\alpha^2 }
        +\frac{(k+\alpha-3\tau)\beta }{3\alpha} \delta(0).
\end{eqnarray}
For localizing the vector zero mode on the positive tension brane, the effective potential $V_{\text{v}}(z)$ should be negative at $z=0$, which  is equivalent to the following condition:
\begin{eqnarray}
 \tau >\frac{1}{3}(k+\alpha).
    \label{vector_norcondition1_1}
\end{eqnarray}
By setting $m=0$ in Eq. (\ref{STbrane_vector_equation}),
we can get the normalized zero mode of the vector field for $~\tau \neq \frac{k+4\alpha}{3}$:
\begin{eqnarray}
 \rho_{0}(z) =\sqrt{\frac{\beta d}
                     {6\alpha\big[(1+\beta z_b)^\frac{d}{3\alpha}-1\big]}}
                      (1+\beta |z|)^\frac{k+\alpha-3\tau}{6\alpha},
\end{eqnarray}
 where $d=k+4\alpha-3\tau$. It is localized on the positive tension brane  when the extra dimension is finite under the condition (\ref{vector_norcondition1_1}). It is easy to see that the vector zero mode can also be localized on the brane when $\tau =0$ if the extra dimension is finite. If the parameter $\tau= \frac{k+4\alpha}{3}$, the normalized vector zero mode is
\begin{eqnarray}
 \rho_{0}(z)=\sqrt{\frac{\beta}{2\ln(1+\beta z_b)}}\frac{1}{\sqrt{1+\beta |z|}}~~
   (\tau= \frac{k+4\alpha}{3}).
\label{STbrane_special_zeromode_vector}
\end{eqnarray}
The vector zero mode (\ref{STbrane_special_zeromode_vector}) can be localized on the positive tension brane only for the case of a finite extra dimension. However, we note here that if the extra dimension is infinite, the zero mode can also be localized on this brane for $\tau >\frac{k+4\alpha}{3}$.

By defining the dimensionless potential
\begin{equation}
\bar{V}_{v}(\bar{z})=\frac{V_{v}(z)}{\beta^{2}},
\end{equation}
Eq.~(\ref{STbrane_vector_equation}) can be rewritten in a dimensionless form,
\begin{equation}
[-\partial^{2}_{\bar{z}}+\bar{V}_{v}(\bar{z})]\rho_{n}(\bar{z})=\bar{m}^{2}_{n}\rho_{n}(\bar{z}),
\end{equation}
where the dimensionless parameters $\bar{z}$ and $\bar{m}_{n}$ are defined as in Eq.~(\ref{scalar dimensionless redefine}).  For simplicity,
we only require that the KK modes satisfy the Neumann boundary condition $\partial_{\bar{z}}(e^{-\epsilon A(\bar{z})}\rho_{n})=0$ at the boundaries $\bar{z}=0$ and $\bar{z}=\bar{z}_b$. Using the boundary condition at $\bar{z}=0$, we get the general solution of Eq. (\ref{STbrane_vector_equation}),
\begin{eqnarray}
 \rho_n(\bar{z})&=&{N} (1+|\bar{z}|)^\frac{1}{2}
              \Big[\text{J}_{P_{\text{v}}} (\bar{m}_{n}(|\bar{z}|+1)) \nonumber\\
            &&+ \mathcal{C}_{\text{v}} ~ \text {Y}_{P_{\text{v}}}(\bar{m}_{n}(|\bar{z}|+1)) \Big],
\end{eqnarray}
where $N$ is the normalization coefficient and
\begin{eqnarray}
  \mathcal{C}_{\text{v}}\equiv -\frac{\text{J}_{P_{\text{v}}-1}(\bar{m}_n)}
  {\text{Y}_{P_{\text{v}}-1}(\bar{m}_n)},~~ {P_{\text{v}}} \equiv {\frac{2\alpha+3\tau-k}{6\alpha}}.\nonumber
\end{eqnarray}
With another boundary condition at $\bar{z}=\bar{z}_b$,  we obtain the spectrum of the vector KK modes determined by the following condition:
\begin{eqnarray}
\mathcal{M}(\bar{m})\equiv&&\text{J}_{P_{\text{v}}-1} (\bar{m}_{n}(\bar{z}_b+1))\nonumber\\
              && +\mathcal{C}_{\text{v}}\text{Y}_{P_{\text{v}}-1} (\bar{m}_{n}(\bar{z}_b+1))=0 .
               \label{STbrane_vector_massive_condition}
\end{eqnarray}

We plot the relation between $\mathcal{M}$ and $\bar{m}$ In Figs. \ref{Spectrum_Vector_Solution1} and \ref{Spectrum_Vector_Solution2} for large and small $\bar{z}_b$, respectively.
Figure \ref{Spectrum_Vector_Solution1a} and Fig. \ref{Spectrum_Vector_Solution1b} imply that the mass of the first excited state increases with the coupling constant of dialton, $\tau$. And Fig. \ref{Spectrum_Vector_Solution1b} and Fig. \ref{Spectrum_Vector_Solution1c} show that in a single cycle the number for the excited states increases with the size of the extra dimension, $\bar{z}_{b}$. In Fig. \ref{Spectrum_Vector_Solution2}, we adjust the size of the extra dimension $\bar{z}_{b}=0.1$ and $\bar{z}_{b}=0.06$ and let $k=-3, \tau=20$ at the same time; the result shows that the excited states do not exist in each cycle and they only emerge after multiple periods. In addition, Fig. \ref{Spectrum_Vector_Solution3} shows that the spectrum interval approaches a constant for the higher excited states, while for the lower excited states it is relatively sparse.
\begin{figure}[htb]
\subfigure[$\tau=1,\bar{z}_{b}=1$] {
\includegraphics[width=0.31\textwidth]{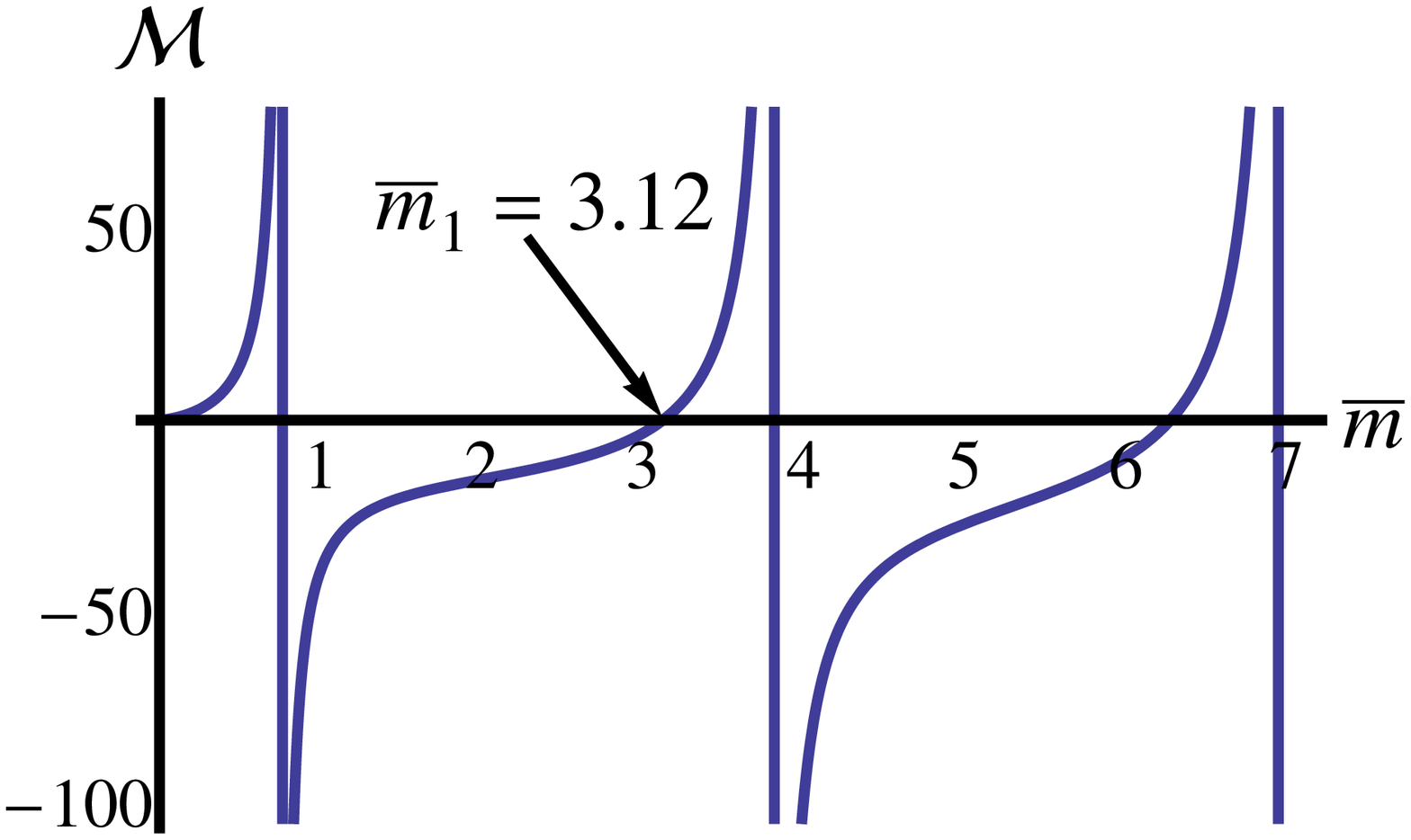}\label{Spectrum_Vector_Solution1a}}
\subfigure[$\tau=20, \bar{z}_{b}=1$] {
\includegraphics[width=0.31\textwidth]{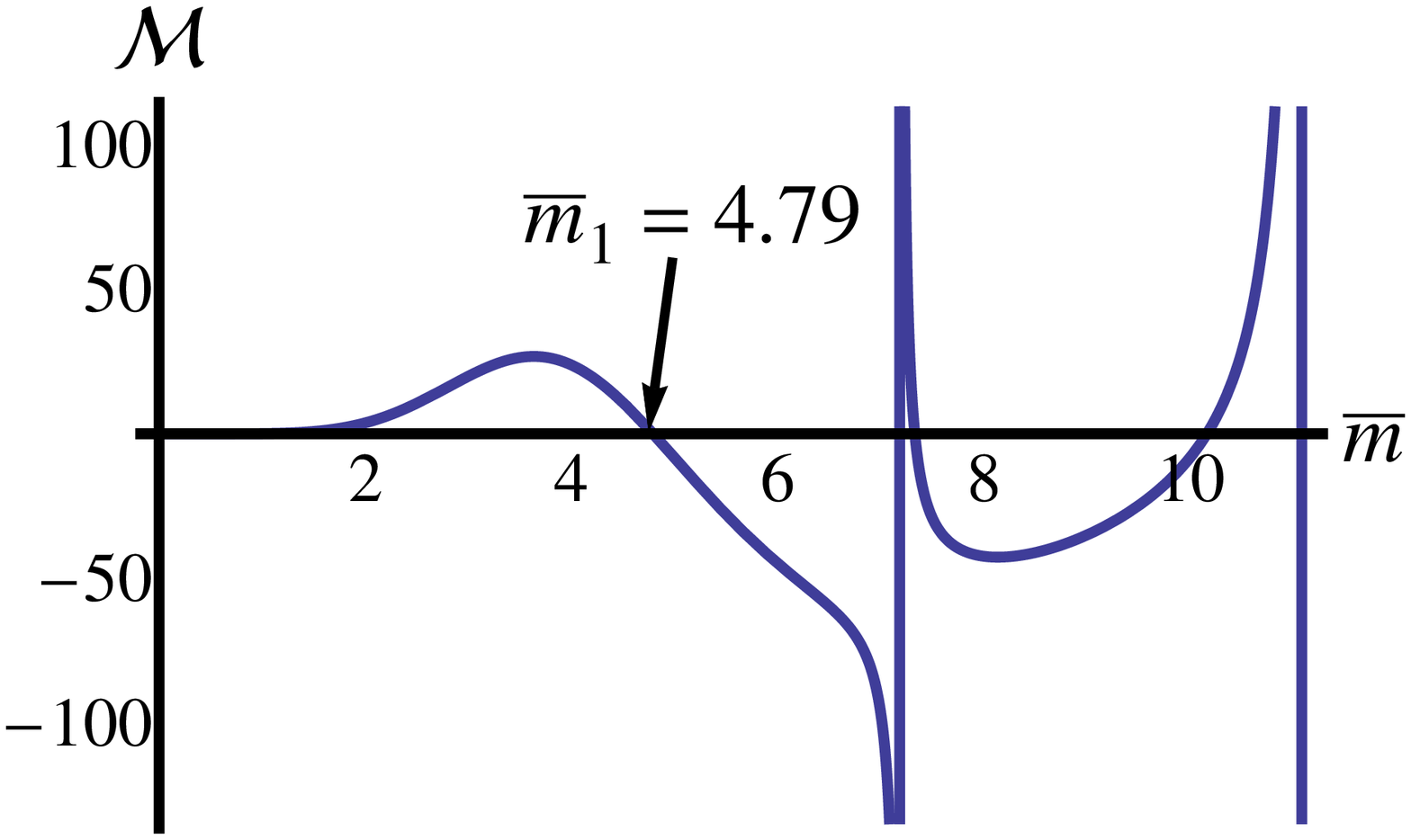}\label{Spectrum_Vector_Solution1b}}
\subfigure[$\tau=20, \bar{z}_{b}=10$] {
\includegraphics[width=0.31\textwidth]{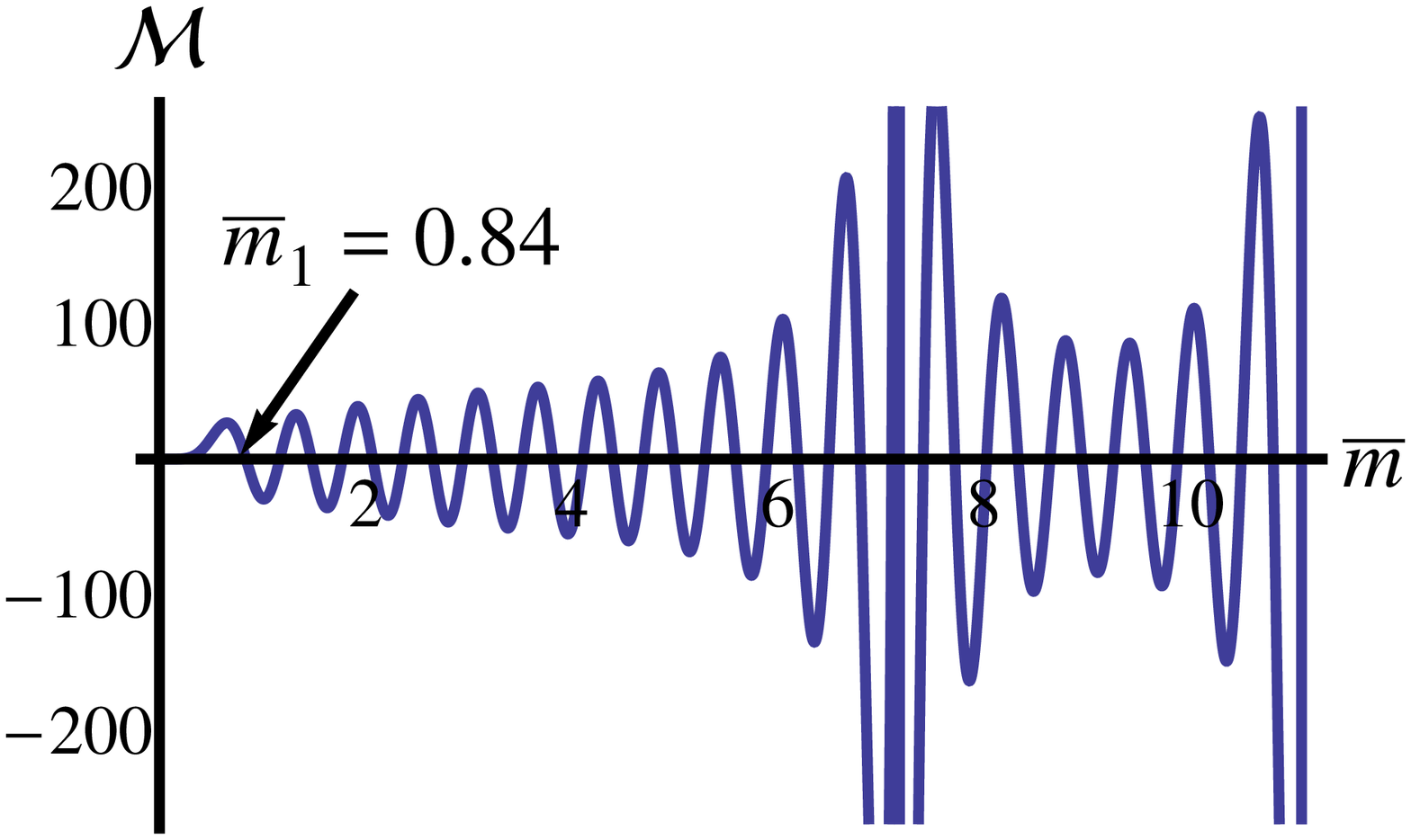}\label{Spectrum_Vector_Solution1c}}
 \caption{The effect of the parameter $\tau$ and the size of the extra dimension $\bar{z}_{b}$ on the massive vector KK modes, and the parameter $k$ is set to $k=-3$.}
 \label{Spectrum_Vector_Solution1}
\end{figure}
\begin{figure}[htb]
\subfigure[$\bar{z}_{b}=0.1$]{
\includegraphics[width=0.43\textwidth]{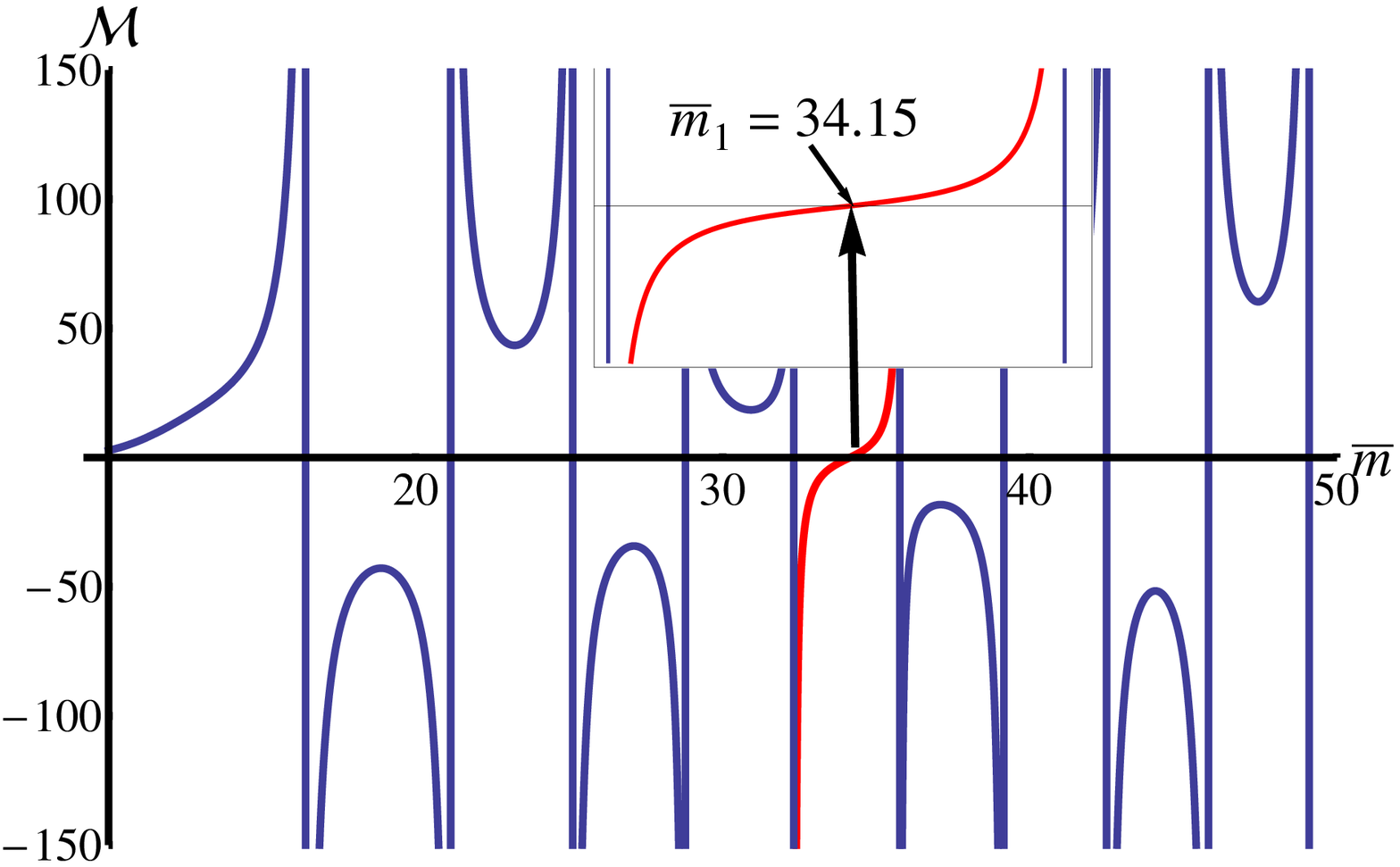}}
\subfigure[$\bar{z}_{b}=0.06$]{
\includegraphics[width=0.43\textwidth]{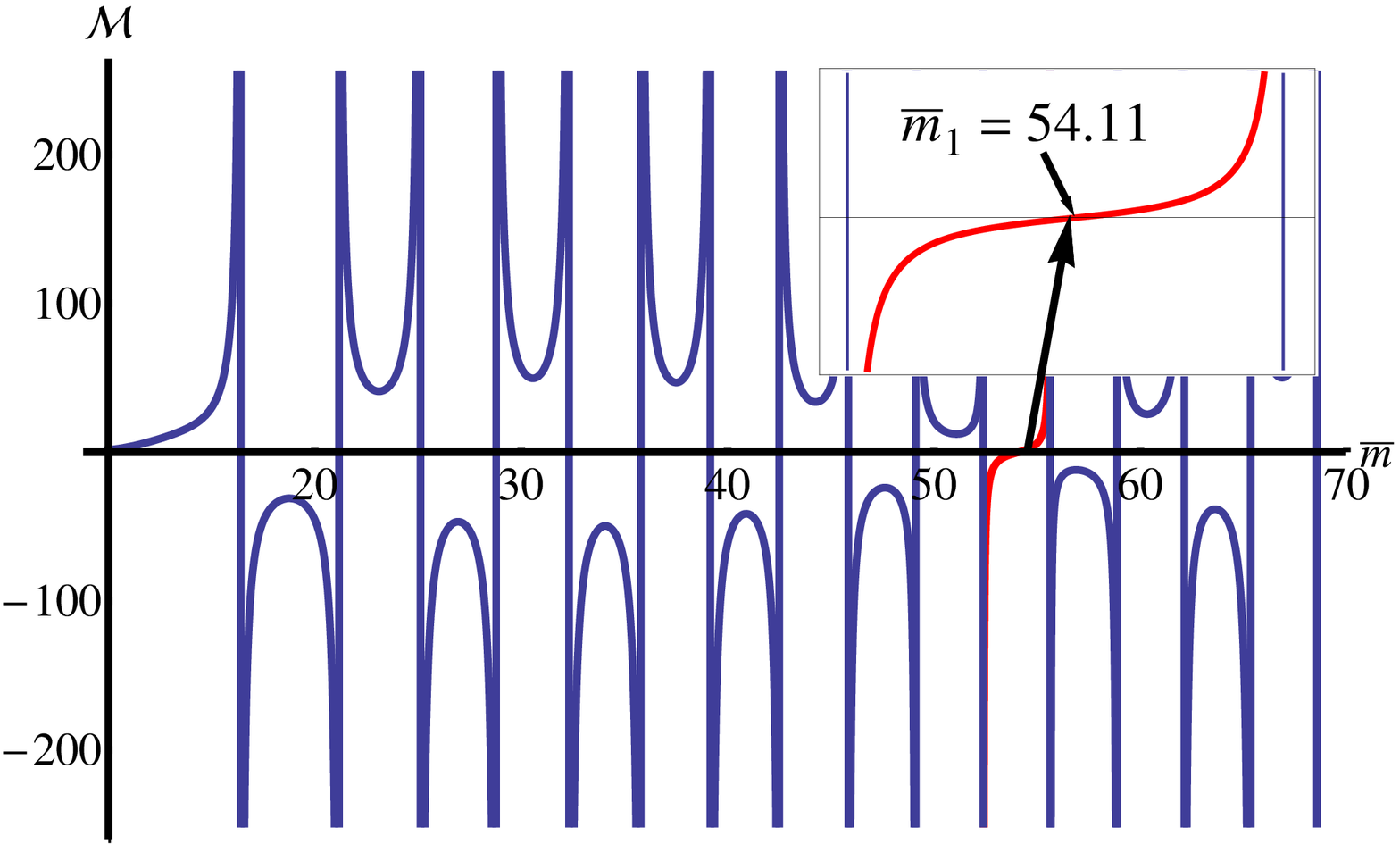}}
 \caption{The effect of the size of the extra dimension $\bar{z}_{b}$ (small $\bar{z}_{b}<1$) on the massive vector KK modes. The parameters are set to $k=-3$, and $\tau=20$.}
 \label{Spectrum_Vector_Solution2}
\end{figure}
\begin{figure}[htb]
\subfigure[$\bar{z}_{b}=50$]{
\includegraphics[width=0.43\textwidth]{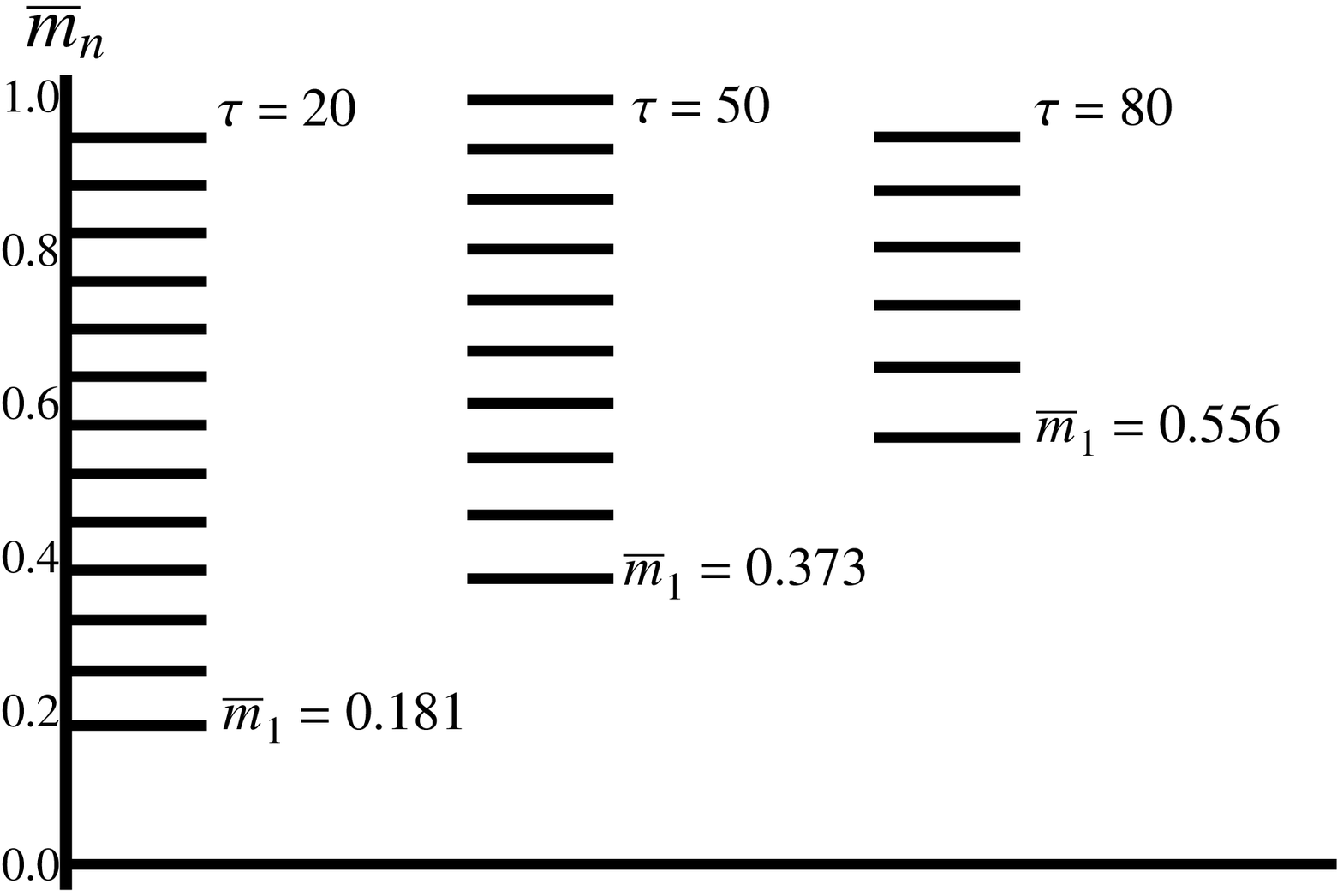}}
\subfigure[$\tau=80$]{
\includegraphics[width=0.43\textwidth]{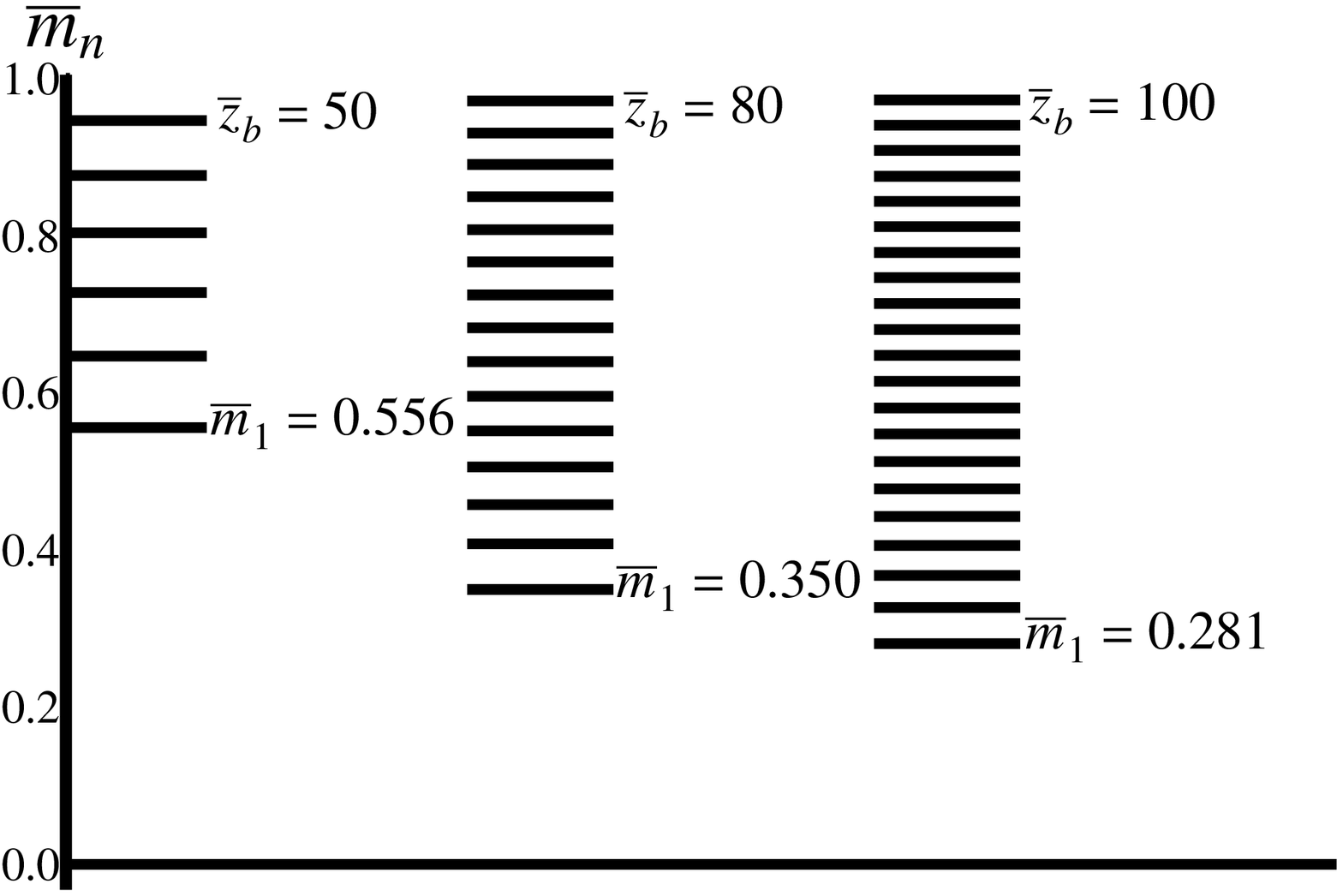}}
 \caption{The mass spectra of the vector KK modes. The parameter $k$ is set to $k=-3$.}
 \label{Spectrum_Vector_Solution3}
\end{figure}

\subsection{Spin-1/2 fermion field}
\label{STbrane_secFermionFields}
 Furthermore, we investigate the localization and mass spectrum of spin-1/2 fermions on the Weyl
 brane. For studying the localization of fermions on thick branes, Refs.~\cite{Liu0708,Volkas2007,
Fuentevilla1412,Cendejas1503,Rocha1606,Brito1609,KoleyCQG2005,D. Bazeia0809,MelfoPRD2006,
Almeida0901,Almeida2009,Davies0705,LBCastro1,LBCastro2} have introduced the Yukawa
coupling. Here, we consider the following five-dimensional Dirac action for a fermion coupled with
the background scalar field $\omega$:
\begin{equation}
S_\frac{1}{2} = \int d^5 x  \sqrt{-g}\left [ \bar \Psi \Gamma^M (\partial_M + \omega_M) \Psi
-\eta \bar \Psi F(\omega) \Psi \right],
\end{equation}
where $\omega_M$ is the spin connection defined as \\$\omega_M=
\frac{1}{4} \omega_M^{\bar{M} \bar{N}} \Gamma_{\bar{M}}
\Gamma_{\bar{N}}$ with $\omega_M ^{\bar{M} \bar{N}}$ given by
\begin{eqnarray}
 \omega_M ^{\bar{M} \bar{N}} &=& \frac{1}{2} {e}^{N \bar{M}}(\partial_M e_N^{~\bar{N}}
                      - \partial_N e_M^{~\bar{N}})\nonumber\\
 &&- \frac{1}{2} {e}^{N\bar{N}}(\partial_M e_N^{~\bar{M}}- \partial_N e_M^{~\bar{M}})  \nn \\
   && - \frac{1}{2} {e}^{P \bar{M}} {e}^{Q \bar{N}} (\partial_P e_{Q
{\bar{R}}} - \partial_Q e_{P {\bar{R}}}) {e}_M^{~\bar{R}}.
\end{eqnarray}
The non-vanishing components of the spin connection $\omega_M$ for
the background metric (\ref{STbrane_linee1}) are
\begin{eqnarray}
  \omega_\mu =\frac{1}{2}(\partial_{z}A) \gamma_\mu \gamma_5
             +\hat{\omega}_\mu, \label{spinConnection}
\end{eqnarray}
where $\hat{\omega}_\mu=\frac{1}{4}
\bar\omega_\mu^{\bar{\mu} \bar{\nu}} \Gamma_{\bar{\mu}}
\Gamma_{\bar{\nu}}$ is the spin connection derived from the metric
$\hat{g}_{\mu\nu}(x)=\hat{e}_{\mu}^{~\bar{\mu}}(x)
\hat{e}_{\nu}^{~\bar{\nu}}(x)\eta_{\bar{\mu}\bar{\nu}}$. Then we
can obtain the equation of motion
\begin{eqnarray}
 \left\{ \gamma^{\mu}(\partial_{\mu}+\hat{\omega}_\mu)
         + \gamma^5\left(\partial_z  +2 \partial_{z}A \right)
         -\eta\; \text{e}^A F(\omega)
 \right \}\Psi =0, \label{DiracEq1}
\end{eqnarray}
where $\gamma^{\mu}(\partial_{\mu}+\hat{\omega}_\mu)$ is the Dirac operator on the brane. For the current case of flat brane, the spin connection on the brane vanishes, i.e., $\hat{\omega}_\mu=0$.

We make a general chiral decomposition,
\begin{equation}
 \Psi= \text{e}^{-2A(z)}\sum_n\Big(\psi_{Ln}(x) f_{Ln}(z)
 +\psi_{Rn}(x) f_{Rn}(z)\Big),
 \label{fengjie1}
\end{equation}
where $\psi_{Ln}(x)=-\gamma^5 \psi_{Ln}(x)$ and $\psi_{Rn}(x)=\gamma^5 \psi_{Rn}(x)$ are the left- and right-chiral components of a four-dimensional Dirac field, respectively. Then we can show that $\psi_{L}(x)$ and $\psi_{R}(x)$ satisfy the four-dimensional massive Dirac equations: $\gamma^{\mu}(\partial_{\mu}+\hat{\omega}_\mu)\psi_{Ln}(x)=m_n\psi_{R_n}(x)$ and $\gamma^{\mu}(\partial_{\mu}+\hat{\omega}_\mu)\psi_{Rn}(x)=m_n\psi_{L_n}(x)$, and the KK modes $f_{Ln}(z)$ and $f_{Rn}(z)$ satisfy the following coupled equations
\begin{subequations}\label{CoupleEq1}
\begin{eqnarray}
 \left[\partial_z
                  + \eta\;\text{e}^A F(\omega) \right]f_{Ln}(z)
  &=&  ~~m_n f_{Rn}(z), \label{CoupleEq1a}  \\
 \left[\partial_z
                  - \eta\;\text{e}^A F(\omega) \right]f_{Rn}(z)
  &=&  - m_n f_{Ln}(z). \label{CoupleEq1b}
\end{eqnarray}
\end{subequations}
From the above coupled equations, we can obtain the Schr\"{o}dinger-like equations for the left- and right-chiral KK modes of the fermion
\begin{eqnarray}
  \big(-\partial^2_z + V_L(z) \big)f_{Ln}
            &=&m_{L_n}^{2} f_{Ln},~~
   \label{SchEqLeftFermion}  \\
  \big(-\partial^2_z + V_R(z) \big)f_{Rn}
            &=&m_{R_n}^{2} f_{Rn},
   \label{SchEqRightFermion}
\end{eqnarray}
where the effective potentials are given by
\begin{subequations}\label{Vfermion}
\begin{eqnarray}
  V_L(z)&=& \big(\eta\text{e}^{A} F(\omega)\big)^2
     - \partial_z \big(\eta\text{e}^{A} F(\omega)\big), \label{VL}\\
  V_R(z)&=&   V_L(z)|_{\eta \rightarrow -\eta}. \label{VR}
\end{eqnarray}\label{LR}
\end{subequations}
In order to get the effective four-dimensional action for a massless fermion and a set of massive Dirac fermions
\begin{eqnarray}
 S_{\frac{1}{2}}=\sum_{n}\int d^4 x~\sqrt{-\hat{g}}
    ~\bar{\psi}_{n} [\gamma^{\mu}(\partial_{\mu}+\hat{\omega}_\mu) -m_{n}]\psi_{n},
\end{eqnarray}
we need to introduce the following orthonormality conditions:
\begin{eqnarray}
 &&\int_{-z_{b}}^{z_{b}} f_{Lm} f_{Ln}dz= \int_{-z_{b}}^{z_{b}} f_{Rm} f_{Rn}dz=\delta_{mn},\nonumber\\
&&\int_{-z_{b}}^{z_{b}} f_{Lm} f_{Rn}dz=0.
 \label{STbrane_orthonormality}
\end{eqnarray}

We set $F(\omega)=\partial_{z}(e^{\upsilon\omega})$ as a simple example.
The explicit expressions of the effective potentials read
\begin{eqnarray} \label{STbrane_VLVRfermion_explicit1}
 V_L(z) &=& \frac{v\eta\beta^2}{\alpha^2}(1+\beta |z|)^{\frac{k-5\alpha-3v}{3\alpha}}
        \Big[ v\eta(1+\beta |z|)^{\frac{k+\alpha-3v}{3\alpha}}\nonumber\\
         &&+\frac{k-2\alpha-3v}{3}\Big] \nonumber\\
        &&+{\frac{2v \eta \beta }{\alpha}}\Big[ \delta(z)- \frac{1}{1+\beta z_b} \delta(z-z_b) \Big], \\
      V_R(z) &=& V_L(z)|_{\eta \rightarrow -\eta}.
\end{eqnarray}
One can easily calculate the values of $V_{L}(z)$ and $V_{R}(z)$  at $z = 0$:
\begin{eqnarray}
 V_L(0)&=&  \frac{v \eta\beta^2}{\alpha^2}
        \left[ v\eta-v-\alpha+\frac{1}{3}(k+\alpha)\right] \nonumber\\
        &&+{\frac{2v \eta \beta }{\alpha}}\delta(0),~\\
        V_R(0) &=& V_L(0)|_{\eta \rightarrow -\eta}.
\end{eqnarray}
It can be shown that the left- and right-chiral fermion zero modes cannot be localized on the positive
tension brane at the same time. For localizing the zero mode of the left-chiral fermion on the positive
tension brane, the effective potential $V_{L}(z)$ should be negative at $z=0$, which requires that $\eta v<0$, since $\beta$ and $\alpha$ are positive.

In the following, we suppose the left-chiral fermion has the zero mode localized on the positive
tension brane. So we only consider the left-chiral fermion (the right-chiral fermion will have the same
mass spectrum with $m_n>0$). The solution of the left-chiral fermion zero mode is
\begin{eqnarray}
 f_{L0}(z)& \propto&
          \exp\left(-\eta \int dz' e^{A(z')}F(\omega)\right).
        \label{STbrane_zeroModefL10}
          \end{eqnarray}
For the case of infinite extra dimension, in order to localize the zero mode $f_{L0}(z)$ on the brane,
the following normalization condition should be satisfied:
\begin{eqnarray}
 \int_{-\infty}^{\infty} dz {f_{L0}}^2(z) \propto
 \int_{-\infty}^{\infty} e^{\left(-2\eta \int^z d\bar{z} e^{A(\bar{z})}F(\bar{\omega})\right)} dz <\infty,
\end{eqnarray}\label{STbrane_eq:Norcz}
which turns out to be $v<\frac{1}{3} \left(k+\alpha\right)~(<0)$ and $\eta >0$. Next, we consider the
case of finite extra dimension, for which we only need $\eta v<0$ in order for the zero mode $f_{L0}$ to
be localized on the positive tension brane. We will investigate the localization and mass spectrum for the
special case $v=\frac{1}{3}(k+\alpha)<0$. The effective potential for the left-chiral fermion KK modes is
reduced to
\begin{eqnarray}
 V_{L}(z)&=&\frac{v\eta\beta^2}{\alpha^2(1+\beta|z|)^2}(v\eta-\alpha)\nonumber\\
 &&+\frac{2v\eta\beta}{\alpha} \Big[\delta(z)-\frac{1}{1+\beta z_b}\delta(z-z_b)\Big].\label{STbrane_old 1}
\end{eqnarray}
Now, in order to transform Eq. (\ref{SchEqLeftFermion}) into a dimensionless one, we
redefine the following dimensionless potential:
\begin{equation}
\bar{V}_{L}(\bar{z})=\frac{V_{L}(z)}{\beta^{2}}.
\end{equation}
We assume the KK modes satisfy the Neumann boundary condition
$\partial_{\bar{z}} (e^{-2A(\bar{z})}f_{Ln})= 0$ at the boundaries $\bar{z}=0$ and $\bar{z}=\bar{z}_b$.
By using the boundary condition at $\bar{z}=0$, we get the general solution of Eq. (\ref{SchEqLeftFermion}),
\begin{eqnarray}
 f_{Ln}(\bar{z})={N} (1+|\bar{z}|)^\frac{1}{2}
            \Big[
                 \text{J}_{P_{\text{f}}} (\bar{z}) +
      \mathcal{C}_{\text{f}} ~ \text {Y}_{P_{\text{f}}}(\bar{z})
            \Big],
\end{eqnarray}
where
\begin{eqnarray}
  \mathcal{C}_{\text{f}}\equiv -\frac{\text{J}_{P_{\text{f}}-1}(\bar{m}_n)}
  {\text{Y}_{P_{\text{f}}-1}(\bar{m}_{n})},~~
   {P_{\text{f}}}
  \equiv {\frac{\sqrt{(-3\alpha+2k\eta+2\alpha\eta)^2}}{6\alpha}}.
\end{eqnarray}
Another boundary condition will give the mass spectrum of the fermion KK modes:
\begin{eqnarray}
\mathcal{M}\equiv(4k+\alpha+6P_{\text{f}}\alpha) B_{P_{\text{f}}}
                             -(4\bar{m}_n\alpha\bar{z}_b) B_{P_{\text{f-1}}}=0,
                             \end{eqnarray}
where $B_{P_{\text{f}}}$ and $B_{P_{\text{f-1}}}$ are defined as follows:
\begin{eqnarray}
B_{P_{\text{f}}}\equiv\text{J}_{P_{\text{f}}} (\bar{m}_{n}(\bar{z}_b+1))
              +\mathcal{C}_{\text{f}}\text{Y}_{P_{\text{f}}}(\bar{m}_{n}(\bar{z}_b+1)), \\
B_{P_{\text{f-1}}}\equiv\text{J}_{P_{\text{f}}-1} (\bar{m}_{n}(\bar{z}_b+1))
          +\mathcal{C}_{\text{f}}\text{Y}_{P_{\text{f}}-1} (\bar{m}_{n}(\bar{z}_b+1)).
\end{eqnarray}
We can obtain the mass spectrum of the left-chiral fermions by numerical calculation, and show
some results in Figs. \ref{Spectrum_fermion_SolutionL1}, \ref{Spectrum_fermion_SolutionL2},
and \ref{Spectrum_fermion_SolutionL3}. Figure \ref{Spectrum_fermion_SolutionL1} shows that the mass
of the first massive mode increases with the value of the Yukawa coupling constant $\eta$, and the
number of the excited states in a single period increases with the size of the extra dimension
$\bar{z}_{b}$. Figure \ref{Spectrum_fermion_SolutionL2} indicates that, for the case of $\bar{z}_b<1$,
the excited states will not appear in every period and they only emerge after multiple periods.
We plot the mass spectrum in Fig. \ref{Spectrum_fermion_SolutionL3} for different parameters.
It can be seen that the spectrum interval approaches a constant for the higher excited states due to
the plane wave behavior of the wave function, while it is relatively sparse for the lower excited states.
\begin{figure}[htb]
\subfigure[$\eta=1,\bar{z}_{b}=1$]{
\includegraphics[width=0.32\textwidth]{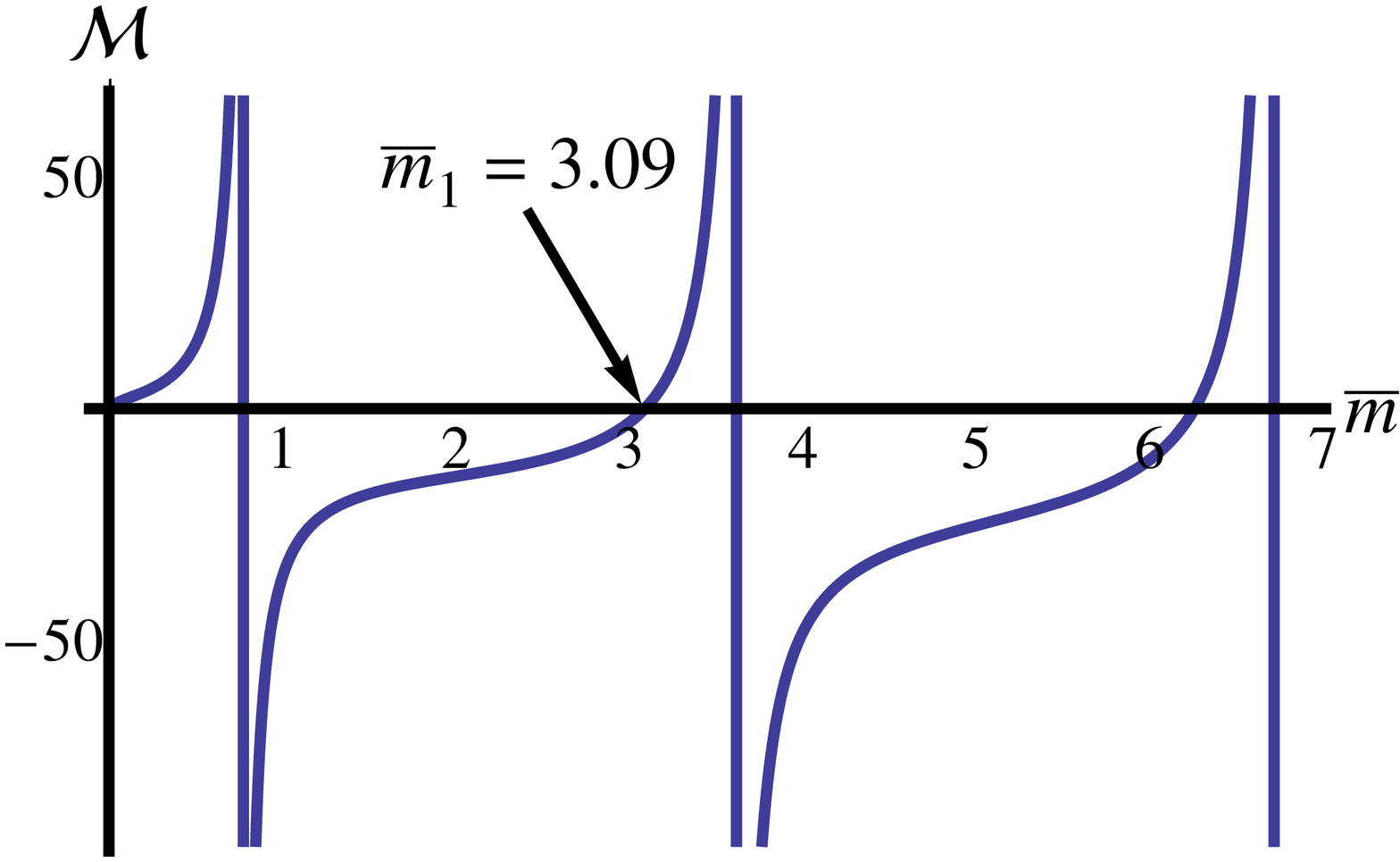}}
\subfigure[$\eta=10,\bar{z}_{b}=1$]{
\includegraphics[width=0.32\textwidth]{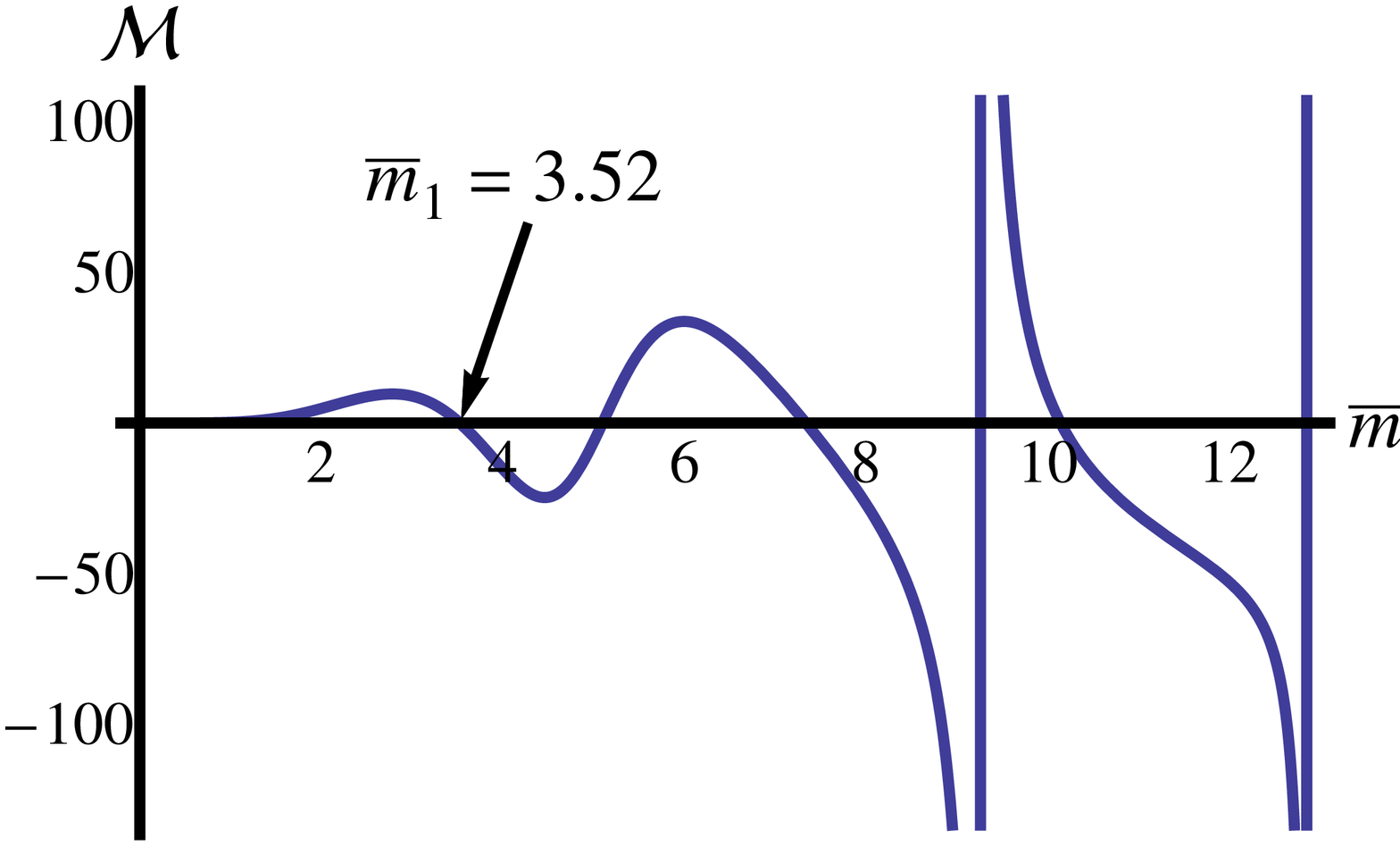}}
\subfigure[$\eta=20,\bar{z}_{b}=5$]{
\includegraphics[width=0.32\textwidth]{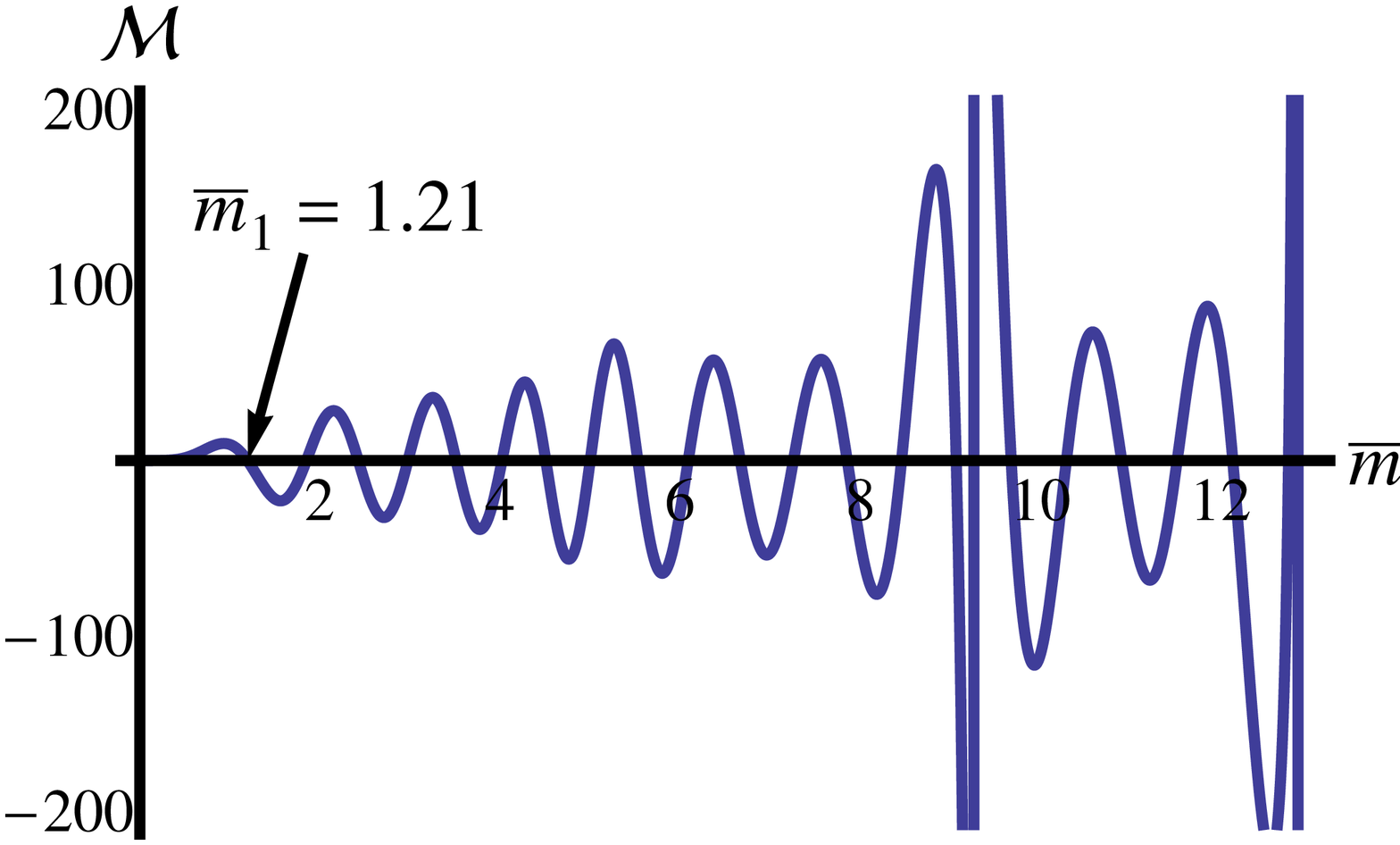}}
 \caption{The effect of the Yukawa coupling constant $\eta$ and the size of the extra dimension
 $\bar{z}_{b}$ on the mass spectrum of the left-chiral fermion KK modes.
 The parameter $k$ is set to $k=-3$.}
\label{Spectrum_fermion_SolutionL1}
\end{figure}
\begin{figure}[htb]
\subfigure[$\bar{z}_{b}=0.1$]{
\includegraphics[width=0.47\textwidth]{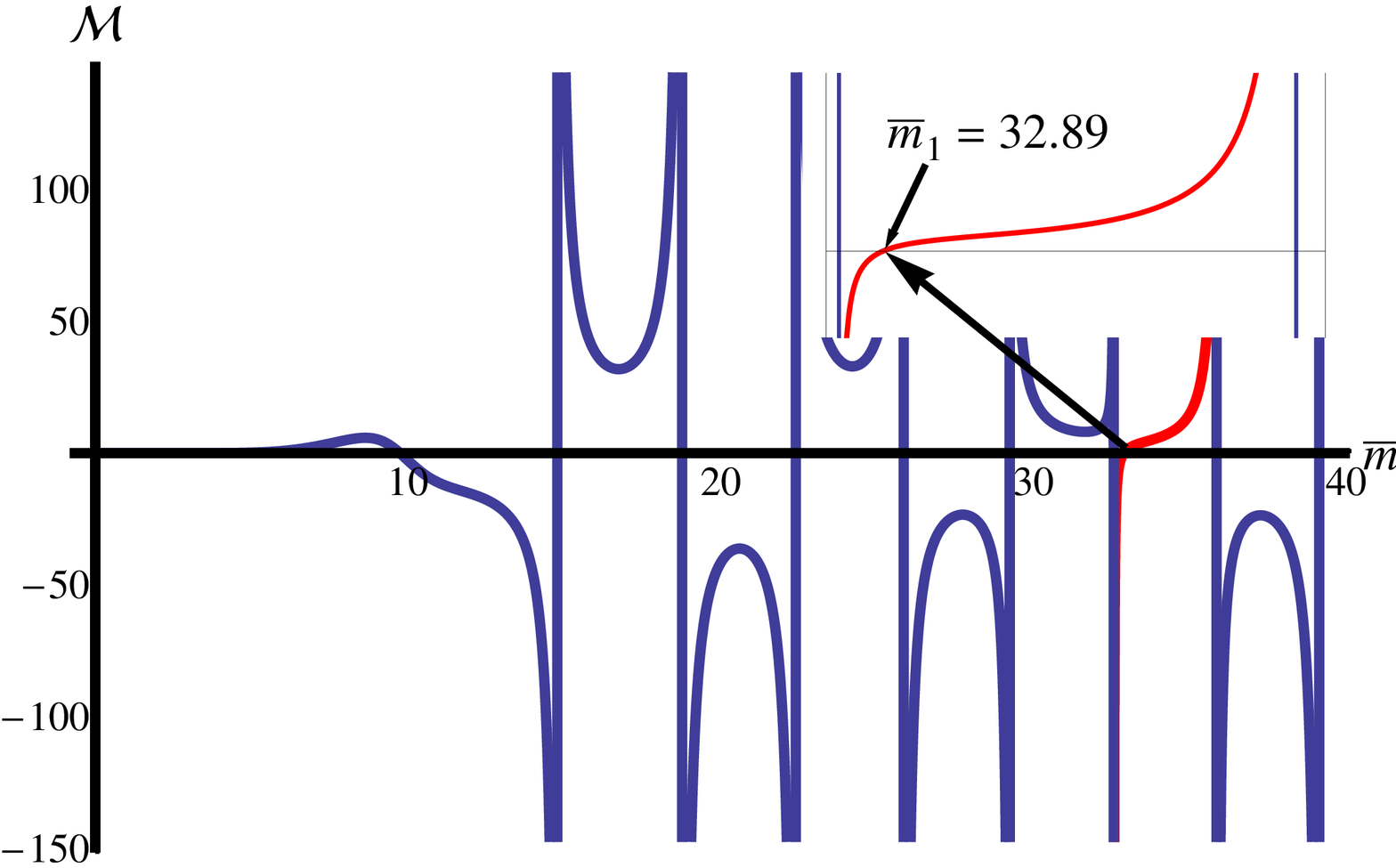}}
\subfigure[$\bar{z}_{b}=0.05$]{
\includegraphics[width=0.47\textwidth]{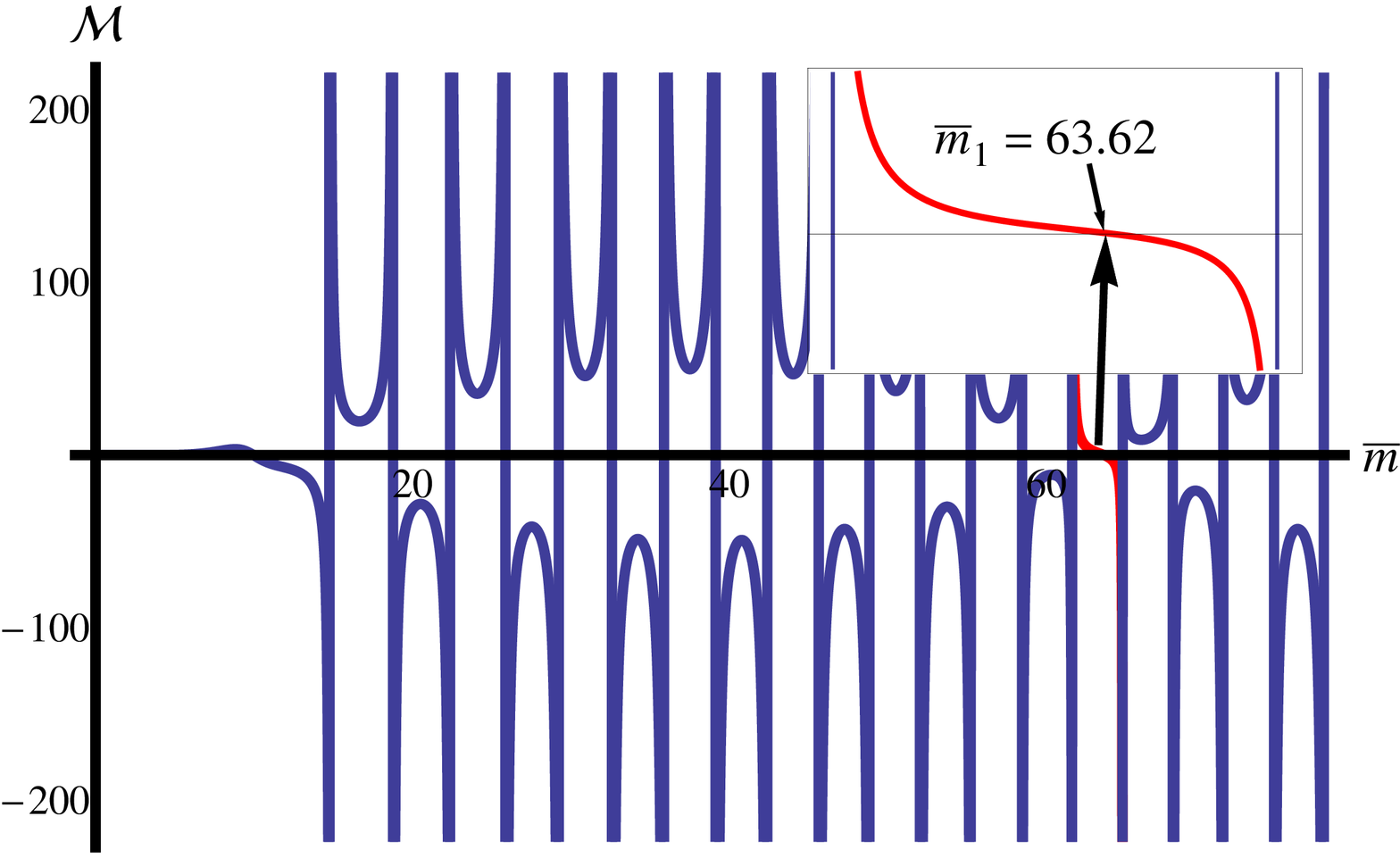}}
 \caption{The effect of the size of the extra dimension $\bar{z}_{b}$ (small $\bar{z}_{b}<1$)
   on the mass spectrum of the left-chiral fermion KK modes.
   The other parameters are set to $k=-3$ and $\eta=40$.}
\label{Spectrum_fermion_SolutionL2}
\end{figure}
\begin{figure}[htb]
\subfigure[$\bar{z}_{b}=30$]{
\includegraphics[width=0.47\textwidth]{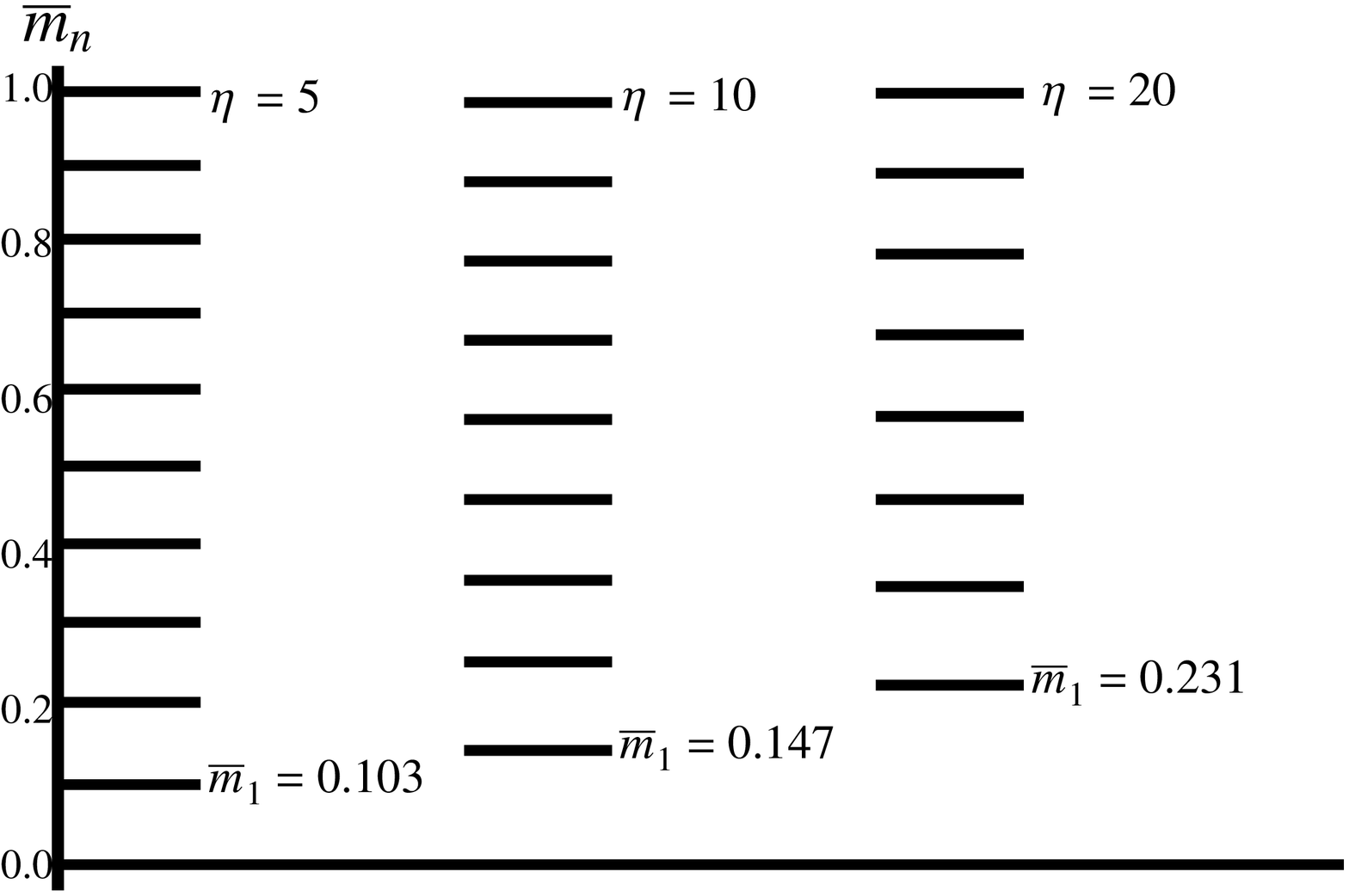}}
\subfigure[$\eta=20$]{
\includegraphics[width=0.47\textwidth]{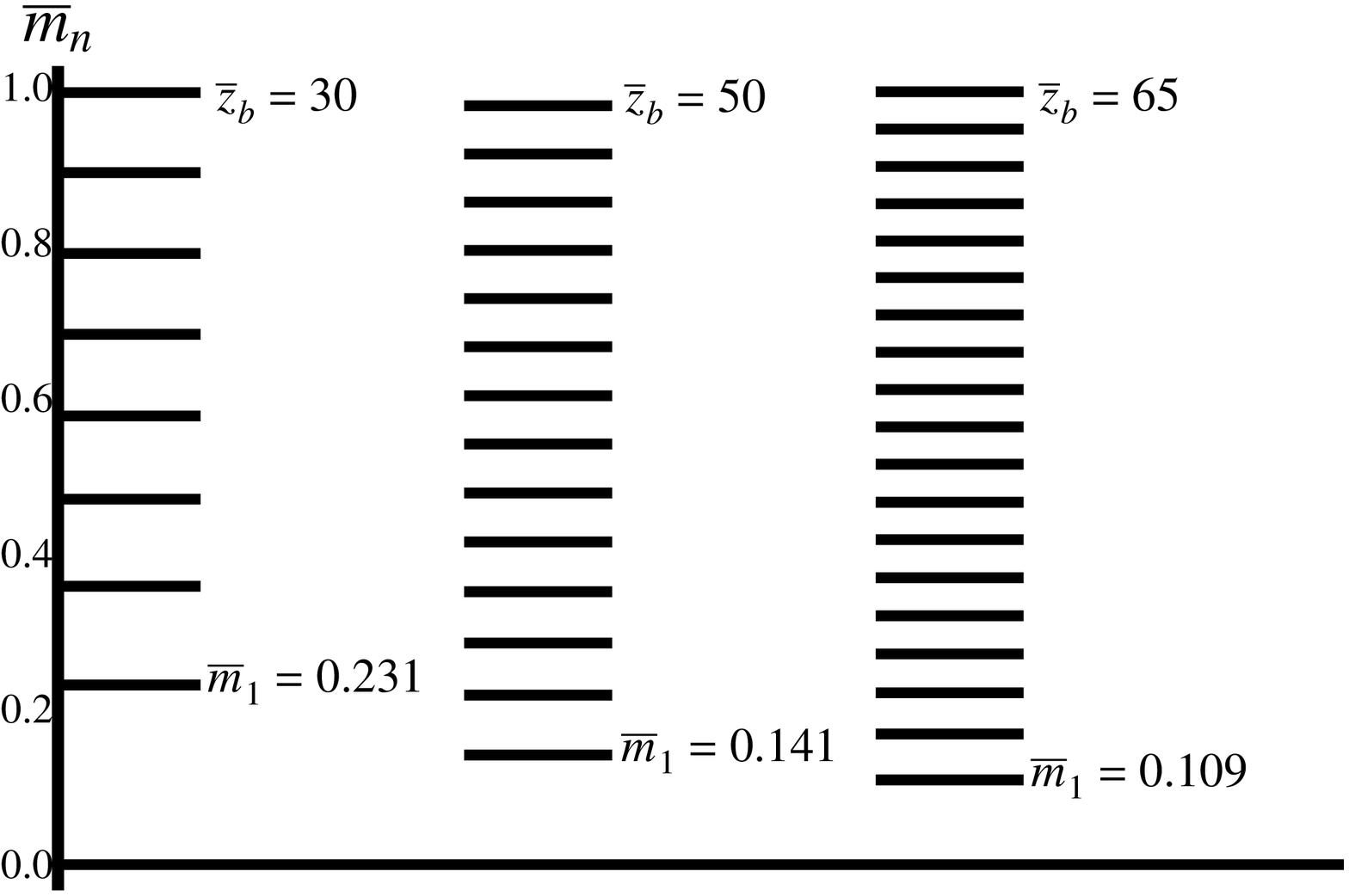}}
 \caption{The mass spectrum of the left-chiral fermion KK modes for different parameters.
 The parameter $k$ is set to $k=-3$ .}
\label{Spectrum_fermion_SolutionL3}
\end{figure}

\section{Correction to Newtonian potential}\label{gravitational perturbations}

Finally, we consider the tensor fluctuations in this model. Now we add a small perturbation
$a^2(z){h}_{\mu\nu}(x,z)$ to the background metric \eqref{STbrane_linee}:
\begin{eqnarray}
ds^2=a^2(z)[(\eta_{\mu\nu}+{h}_{\mu\nu}(x,z))dx^{\mu}dx^{\nu}+dz^2]. \label{Perturbational_Metic}
\end{eqnarray}
Following Ref. \cite{dewolfe}, we suppose the tensor perturbation ${h}_{\mu \nu}$ satisfies the
transverse-traceless (TT) condition:
${h}^{\mu}_{~\mu}=\partial^\nu {h}_{\mu \nu}=0$.
The equation of motion for ${h}_{\mu\nu}$ is given by \cite{Yang:2011pd}
\begin{eqnarray}
{h}''_{\mu\nu}+3\frac{a'}{a}{h}_{\mu\nu}'+k\omega'{h}'_{\mu\nu}+\Box^{(4)}{{h}_{\mu\nu}}=0.
\label{Fluctuation_Eq}
\end{eqnarray}
Furthermore,  ${h}_{\mu\nu}$ can be decomposed in the form
\begin{eqnarray}
{h}_{\mu\nu}(x,z)=\sum_n \varepsilon^{(n)}_{\mu\nu}(x)H^{-\frac{3}{2}}(z)\Psi_n(z)\label{Decomposition},
\end{eqnarray}
where $H(z)=(1+\beta |z|)^{1/3}$. The four-dimensional mass $m_{n}$ of a KK excitation is defined by the
 Klein--Gordon equation $\Box^{(4)}{\varepsilon^{(n)}_{\mu\nu}(x)}=m_n^2 \varepsilon^{(n)}_{\mu\nu}(x)$.
 We obtain a Schr$\ddot{\text{o}}$dinger-like equation  from Eq. (\ref{Fluctuation_Eq}),
\begin{eqnarray}
-\Psi''_n (z)+V(z)\Psi_n(z)=m_{n}^2\Psi_n(z),\label{Schrodinger_Eq}
\end{eqnarray}
where the effective potential $V(z)$ is given by
{\begin{eqnarray}
V(z)=\frac{3}{2}\frac{H''}{H}+\frac{3}{4}\frac{H'^2}{H^2}
= \frac{-\beta ^2} {4( {1 + \beta |z| } )^2}
            .\label{Effective_Potential}
\end{eqnarray}}
The general solution of Eq. (\ref{Schrodinger_Eq}) is a linear combination of the Bessel functions,
\begin{eqnarray}
 \Psi_{n}(z)&=&N_{n}(\frac{1}{\beta}+ |z|)^{\frac{1}{2}}\Big[J_{0}\Big(m_{n}(|z|+\frac{1}{\beta})\Big)\nonumber\\
 &&+ \alpha_n Y_{0}\Big(m_{n}(|z|+\frac{1}{\beta})\Big)\Big].
 \label{equaction solution}
\end{eqnarray}
Imposing the Neumann boundary
condition $\partial_{z}{h}_{\mu\nu}(x,z)=0$ at the boundaries $z=0$ and $z=z_{b}$,
we get \\
$\alpha_{n}=-J_{1}(\frac{m_{n}}{\beta})/Y_{1}(\frac{m_{n}}{\beta})$ and
\begin{equation}
\frac{J_{1}(m_{n}(z_{b}+\frac{1}{\beta}))}{J_{1}(\frac{m_{n}}{\beta})}
=\frac{Y_{1}(m_{n}(z_{b}+\frac{1}{\beta}))}{Y_{1}(\frac{m_{n}}{\beta})}.
\label{graviton mass spectrum}
\end{equation}
The solution of the graviton zero mode $\Psi_{0}(z)$ is
\begin{equation}
\Psi_{0}(z)=N_{0}(\frac{1}{\beta}+ |z|)^{\frac{1}{2}},
\end{equation}
which is localized on the negative tension brane for finite $z_b$. Note that the graviton zero mode cannot
be normalized anymore when the extra dimension is infinite, which is
very different from the RS1 model. The massless zero mode sector of the action (\ref{STbrane_action}) leads to 
us the effective four-dimensional gravitational theory,
\begin{eqnarray}
S_{5}\supset&&\frac{M^{3}_{*}}{2}\int_{M^{W}_{5}}d^{5}x\sqrt{|g|}e^{k\omega}{R} \nonumber\\
               \supset&&\frac{M^{3}_{*}}{2}\int^{z_{b}}_{-z_{b}}dza^{3}(z)e^{k\omega}
                                                    \int_{M_{4}}d^{4}x\sqrt{|g^{(4)}|}\hat{R}^{(4)},
\label{effective }
\end{eqnarray}
where $\hat R^{(4)}$ is the four-dimensional Riemannian Ricci scalar made out of
$g^{(4)}_{\mu \nu}=\eta_{\mu\nu}+h^{(0)}_{\mu\nu}(x)$. Thus the effective four-dimensional Planck
scale $M_{Pl}$ is \cite{Yang:2011pd}
\begin{equation}
M^2_{Pl}=M_{*}^3\int_{-z_b}^{z_b}dz a^3(z)e^{k\omega}=M^3_{*}(2z_b+\beta z_b^2).
\label{Relation_M4_M5}
\end{equation}

For the light modes ($m_{n}/\beta\ll1$), the solution (\ref{equaction solution}) in the long-range limit
$\alpha_{n}\approx\frac{\pi m_{n}^{2}}{4\beta^{2}}\ll1$ becomes
\begin{equation}
\Psi_{n}(z)\approx N_{n}(\frac{1}{\beta}+|z|)^{\frac{1}{2}}\Big[J_{0}(m_{n}(|z|+\frac{1}{\beta}))\Big].
\label{equation solution1}
\end{equation}
Correspondingly, the spectrum of KK gravitons is \cite{Yang:2011pd}
\begin{eqnarray}
  m_{n}(z) = \frac{x_{n}}{z_b + \frac{1}{\beta}}\approx\frac{x_{n}}{z_{b}},
\end{eqnarray}
where $x_n$ satisfies $J_1{(x_n)}=0$, and $x_1$=3.83, $x_2$=7.02,
$x_3$=10.17, $\cdots$.  The normalization constant $N_ n$ is determined by \cite{guobin Add}
\begin{equation}
1=\int_{-z_{b}}^{z_{b}}\Psi_{n}^{2}dz\approx2N^{2}_{n}\int_{0}^{z_{b}}z J_{0}(m_{n}z)^{2}dz
\approx N_{n}^{2}z_{b}^{2}J_{0}^{2}(m_{n}z_{b}).
\label{normalization N}
\end{equation}
Imposing the approximation $J _0 (x) \approx\sqrt{\frac{2}{\pi x}}\cos(x-\frac{1}{4}\pi)$ and the approximate
formula of the zero point of $J_{1} (x_n )$, $ x_n\approx(n+\frac{1}{4})\pi$, we get
$N_ n\approx\sqrt{ \frac{\pi x_n}{2z_{b}^{2}}}$.
Then the normalized KK modes are
\begin{equation}
\Psi_{n}(z)\approx\sqrt{\frac{\pi x_n}{2z_{b}^{2}}}
(\frac{1}{\beta}+|z|)^{\frac{1}{2}} J_{0}(m_{n}(|z|+\frac{1}{\beta})).
\label{equation solution2}
\end{equation}

For simplicity, the effective gravitational potential between two point-like sources with mass $M_1$ and
$M_2$  is obtained by the exchange of the graviton zero mode and massive KK modes, and it can be
expressed as \cite{Randall2,Y.shirman}
\begin{equation}
U(r)=-\frac{M_{1}M_{2}}{M_{Pl}^{2}}~\frac{1}{r}-\frac{M_{1}M_{2}}
{M_{*}^{3}}\sum^{\infty}_{n=1}\frac{e^{-m_{n}r}}{r}~|\Psi_{n}(0)|^{2},
\label{Nweton potention1}
\end{equation}
where the four-dimensional Newtonian potential is caused by the zero mode, and the correction term is
produced by the massive KK modes. Combining Eqs. (\ref{Relation_M4_M5}), (\ref{equation solution2}),
and (\ref{Nweton potention1}), we can get the correction term to the Newtonian potential:
\begin{equation}
\Delta U(r)=-\frac{M_{1}M_{2}}{M^{2}_{Pl}}~\sum^{\infty}_{n=1}~\frac{\pi x_{n}}{2z_{b}\beta}
(2+\beta z_{b})~\frac{e^{-x_{n}\frac{r}{z_{b}}}}{r}.
\label{Nweton potention2}
\end{equation}
For $r\gg z_{b}$ , the summation term tends to $e ^{-x_{n} r/z_{b}}\approx 0$, and the correction can be
ignored. For the case of $r\ll z_{b}$ , the correction term cannot be ignored and it can be calculated as
follows:
\begin{equation}
\Delta U(r) \approx~-\frac{M_{1}M_{2}}{M^{2}_{Pl}}~\frac{z_{b}^{2}}{r^{3}}.
\label{Nweton potention3}
\end{equation}
It shows that, for $r\ll z_{b}$, the Newtonian potential is corrected by a $1/r^{3}$ term, which is the leading term. For $r\gg z_{b}$, the correction can be ignored, and we recover the Newtonian potential $U(r) \propto1/r$.
According to the gravitational experiments of the correction to the Newtonian potential~\cite{Eingorn11,Iorio11,Linares13,Luo2016}, we know that the size of the extra dimension, $z_{b}$, should not be larger than
the micron scale.

\section{Conclusion and summary}\label{STbrane_secConclusion}
To summarize, we have investigated the localization and mass spectra of various bulk matter fields on the Weyl brane. We first gave a brief review of the thin brane arising from a five-dimensional Weyl integrable spacetime. Then, we investigated localization of the zero modes for various bulk matter fields (i.e., scalar, vector, and
 fermion fields) on the brane, and we got the mass spectra of the fields by redefining some dimensionless
 parameters. We also considered the correction to the four-dimensional Newtonian potential from the
 massive KK gravitons.

It was found that the zero modes of various bulk matter fields can be localized on the positive tension
brane under some conditions, which are collected in Table. \ref{table_localization_condition}. It can be seen
 that the localization conditions for the case of infinite extra dimension are stronger than the case of a finite
 extra dimension. When the extra dimension is finite, the scalar and vector zero modes can be localized on the
 positive tension brane even if there is no interaction with the background scalar field (i.e., $\lambda=0$
 and $\tau=0$).

For the scalar field, we considered two types of couplings with the background scalar field. For the case of the scalar--dilaton coupling, we found that the mass of the first massive scalar KK mode increases and decreases with the dimensionless coupling constant $\lambda$ and the size of the extra dimension $\bar{z}_{b}$, respectively, and the gap of the mass spectrum decreases with $\bar{z}_{b}$. When the dimensionless size
$\bar{z}_{b}>1$, the number of the excited states in a single period, shown in Fig.~\ref{STbrane_fig_Spectrum_Scalar2}, increases with $\bar{z}_{b}$. When
$\bar{z}_{b}<1$, the excited states do not appear in each period, they will emerge after several periods.
For the case of the Higgs potential coupling, we fixed the parameter $\bar{\theta}$ and chose the proper value $\bar{u}$, to ensure the zero mode can be localized on the brane. In order to avoid negative $
\bar{m}^{2}_{n}$, the mass parameter $\bar{u}$ in the  Higgs potential should have an upper limit
when the other parameters are fixed.

For the vector field, it was shown that the mass of the first massive vector KK mode increases with the coupling constant $\tau$, and the gap of the mass spectrum decreases with the size of the extra dimension.

For the fermion field, we introduced the usual Yukawa coupling with $F(\omega)=\partial_{z} e^{\upsilon\omega}$, and we found that the left-chiral fermion zero mode can be localized on the positive tension brane at some conditions. We calculated the mass spectrum of the fermion KK modes for
$v=\frac{1}{3}(k+\alpha)$ as an example. The mass of the first excited state increases with the Yukawa coupling constant $\eta$. The size of the extra dimension can also affect the mass gap in the same way as the cases of the scalar and vector fields. The spectra of various bulk matter fields show the same phenomenon: that the spectrum interval approaches a constant for the higher excited states, while it is
relatively sparse for the lower excited states.

For the TT tensor perturbation of the gravitational field, the zero mode is localized on the negative tension brane for the finite extra dimension, and cannot be normalized anymore when the extra dimension is infinite. For the case of the finite extra dimension, the gravity zero mode gives the Newtonian potential, while the gravity massive KK modes will give a correction to the Newtonian potential by a $1/r^{3}$ term when $r \ll z_b$.


\begin{table*}[!htb]
\begin{center}
\caption{Localization conditions for the zero modes of various bulk matter fields. Here $z_b$ is
the size of the extra dimension.}
\label{table_localization_condition}

\centering
\begin{tabular}{|c|c|c|c|}
  \hline
  Bulk matter & Lagrangian & $~~~~z_b~~~$ & ~condition~   \\\hline \hline
  \multicolumn{1}{|c|}{\multirow {2}{*}{Scalar}}
  &\multicolumn{1}{|c|}{\multirow {2}{*}{$\mathcal{L}_0= -\frac{1}{2} e^{\lambda\omega} \partial_M \Phi \partial^M \Phi$}}
           &  finite
           & $\lambda >k+\alpha$\\
           \cline{3-4}
      &  &  infinite
           & $\lambda >k+2\alpha$ \\\hline
  \multicolumn{1}{|c|}{\multirow {2}{*}{Vector}}
  &\multicolumn{1}{|c|}{\multirow {2}{*}{$\mathcal{L}_1= -\frac{1}{4}  e^{\tau\omega} F_{MN}F^{MN}$}}
           &  finite
           & $\tau >\frac{1}{3}(k+\alpha)$ \\
           \cline{3-4}
      &  &  infinite
           & $\tau >\frac{1}{3}(k+4\alpha)$ \\\hline
  \multicolumn{1}{|c|}{\multirow {2}{*}{Fermion}}
  &\multicolumn{1}{|c|}{\multirow {2}{*}{$\mathcal{L}_{1/2}= \bar \Psi \Gamma^M (\partial_M + \omega_M) \Psi-\eta \bar \Psi F(\omega) \Psi$}}
           &  finite
           & $\eta v <0$ \\
           \cline{3-4}
      &  &  infinite
           & $\eta>0, v<\frac{k+\alpha}{3}$ \\
  \hline
\end{tabular}
\end{center}
\end{table*}

\section*{Acknowledgements}
This work was supported by the National Natural Science Foundation of China (Grants Nos. 11522541 and 11375075) and the Fundamental Research Funds for the Central Universities
 ((Grants No. lzujbky-2016-k04)).


\end{document}